\documentclass[preprint,superscriptaddress,nofootinbib]{revtex4-1}

\usepackage[english]{babel}

\usepackage[letterpaper,top=2cm,bottom=2cm,left=3cm,right=3cm,marginparwidth=1.75cm]{geometry}

\usepackage{amsmath}
\usepackage{physics}
\usepackage{graphicx}
\usepackage{comment}
\usepackage{subcaption}
\usepackage[colorlinks=true, allcolors=blue]{hyperref}

\begin{document}
\title{Asymptotic behavior of the Fourier coefficients and the analytic structure of the QCD equation of state}

\author{Miles Bryant}
\affiliation{Department of Physics, North Carolina State University,
Raleigh, NC 27695}
\author{Christian Schmidt} 
\affiliation{Fakultät für Physik, Universität Bielefeld, D-33615 Bielefeld, Germany}
\author{Vladimir V. Skokov}
\affiliation{Department of Physics, North Carolina State University,
Raleigh, NC 27695}

\begin{abstract}
In this paper we study  the universal properties of the baryon chemical potential Fourier coefficients in Quantum Chromodynamics.     
We show that by following a well-defined strategy,  the  Fourier coefficients can be used to locate Yang-Lee edge singularities  associated with chiral phase transition (and by extension with the Roberge-Weiss)  in the complex chemical potential plane. We comment on the viability of performing this analysis using lattice QCD data.   
\end{abstract}
\maketitle

\section{Introduction}

Due to the sign problem, the lattice Monte-Carlo methods cannot directly probe QCD thermodynamics at non-zero real values of the baryon chemical potential. One thus has to utilize indirect approaches~\footnote{Despite the recent developments in complex Langevin and contour deformation methods (see e.g. Ref.~\cite{Berger:2019odf, Attanasio:2020spv, Alexandru:2020wrj}), direct calculations at non-zero real baryon chemical potential are still not practical.}. One of them is the evaluation of QCD thermodynamics at imaginary chemical potential with the goal of either analytic continuation~\cite{deForcrand:2002hgr, DElia:2002tig}, to the real axis or studying the analytic structure of the  QCD equation of state in the complex chemical potential plane \cite{Dimopoulos:2021vrk}. In this paper, we follow the latter path.  

Consider the QCD equation of state at zero chemical potential above the possible chiral critical point but below the Roberge-Weiss (RW) phase transition~\cite{Roberge:1986mm,Bornyakov:2022blw}, that is, at temperatures in the range $(T_c, T_{\rm RW})$.  For physical quark masses, the equation of state in this regime shows a smooth non-critical transition from hadrons to quark and gluon degrees of freedom. This transition is colloquially referred to as the ``QCD crossover''.  The apparent smoothness of the QCD crossover obscures non-trivial critical behavior at complex values of the baryon chemical potential, where the remnants of the critical point reside, known by the name of Yang-Lee edge singularities~\cite{Kortman:1971zz,Fisher:1978pf}.   These singularities are continuously connected to the associated critical points~\cite{Lee:1952ig,Yang:1952be,Itzykson:1983gb,Stephanov:2006dn}:   when two YLE singularities merge and pinch a  physical axis of the corresponding thermodynamic variable (for the case of the chiral critical point, the baryon chemical potential),   the critical point with corresponding critical scaling emerges. Thus, locating and especially tracking the Yang-Lee edge singularities as a function of temperature may reveal the existence and the location of the QCD critical point.\footnote{Mere presence of the YLE singularities does not necessitate the existence of a critical point at {\it finite}  temperature. For instance, there are YLEs in the classic one-dimensional Ising model, but no critical point for $T>0$.}   These singularities can be treated as standard critical points, with the exception that in contradistinction to standard criticality, there is only one (not two) relevant variable, and thus, there is only one independent critical exponent: the edge critical exponent $\sigma_{\rm YLE} = \frac{1}{\delta_{\rm YLE}}$. The bootstrap approach provides the most accurate estimate for the value of the critical exponent $\sigma_{\rm YLE} = 0.085(1) $~\cite{Gliozzi:2014jsa}. It is independent of the symmetry of the underlying universality class.

In this paper, using the universal properties of the YLE singularities, we predict the asymptotic behavior of the Fourier coefficients of the baryon chemical potential. 
In contrast to the study in Ref.~\cite{Almasi:2019bvl}, our analysis is focused on the cross-over region $T_c<T<T_{\rm RW}$. 
Using this asymptotic result as a fitting prior allows us to locate the YLE singularities in a low-energy QCD model. The model calculations are done in both mean-field approximation (in the mean-field, $\sigma_{\rm YLE}=\frac12$) and by accounting for fluctuations in the local potential approximation of the Functional Renormalization Group approach (in this case, the resulting critical exponent $\sigma_{\rm YLE}=\frac15$ approximate better the actual value).

The outline of the paper is as follows. In Section~\ref{Sec:FC}, we discuss the Fourier coefficients and their asymptotic behavior based on the universal properties. In Section~\ref{Sec:MR}, we apply our analysis to the low energy model and demonstrate that even with a limited number of the Fourier coefficients, the position of the closest YLE singularity to the imaginary axis can be accurately extracted.  We conclude with Section~\ref{Sec:Con}. Moreover, in Appendices, we discuss the finite size effects, the required precision of calculating the baryon number at imaginary values of the chemical potential to extract $k$-th order Fourier coefficient, and, finally, we comment on the numerical quadrature method in computing the Fourier coefficients on the available lattice QCD data.

\section{Fourier coefficients}
\label{Sec:FC}

The QCD partition function along with thermodynamic quantities derivable from it, is periodic in baryon chemical potential $\hat \mu = \mu/T$: 
\begin{align}
 Z\left(\hat \mu + 2\pi i \right) = Z\left(\hat \mu\right)   \,. 
\end{align}
This is why analyzing the data obtained for purely imaginary values of baryon chemical potential is natural in terms of the corresponding Fourier coefficients~\cite{Roberge:1986mm,Vovchenko:2017gkg,Vovchenko:2017xad,Almasi:2018lok,Bornyakov:2016wld,Bornyakov:2022blw,Schmidt:2023jcv}. 
Specifically, from a lattice QCD perspective, it is convenient to compute the Fourier transformation of the baryon number density $n_B = V^{-1} \partial_{\bar \mu} \ln Z(\bar \mu)$:
\begin{align}
    \label{Eq:bk}
    b_k = \frac{1}{i \pi} \int_{-\pi}^{\pi} d\theta  \,  \hat n_B (\hat \mu = i \theta ) 
    \sin \left(k \theta \right),
\end{align}
where we explicitly took into account the symmetry property of the baryon density $ n_B(\hat \mu) = -n_B(-\hat \mu)$ and introduced $\hat n_B = n_B/T^3$.

\begin{figure}
    \centering
    \includegraphics{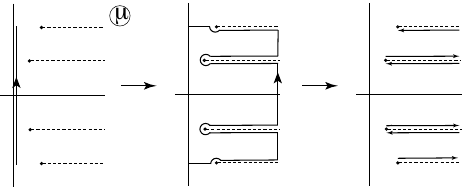}
    \caption{Complex chemical potential plane with the chiral and RW YLE. The integration path along the imaginary chemical potential axis can be deformed to the integration around the branch point singularities and the associated cuts (Stokes lines). }
    \label{fig:countour}
\end{figure}

In actual lattice QCD, the coefficients $b_k$ suffer not only from  statistical error originating from the Monte-Carlo nature of the simulation\footnote{Here, we assume that the continuum and thermodynamic limits are properly taken.} but also from the discretization errors of the numerical integration in Eq.~\eqref{Eq:bk}. In our analysis, we assume that an effort is taken to minimize both types of errors and ignore them. We review the effect of finite volume Appendix~\ref{sec:AppA} and effect of  the statistical errors in Appendix~\ref{Sec:AppB}. We propose a specific numerical quadrature for the evaluation of Eq.~(\ref{Eq:bk}) in Appendix~\ref{Sec:AppC}.


The question is, then, what can we learn from analyzing the Fourier coefficients?  The answer is quite a bit. Consider the integral~\eqref{Eq:bk} in the complex plane, as illustrated in Fig.~\ref{fig:countour}. In this figure, we took into account the presence of the YLE singularities associated with the chiral and Roberge-Weiss phase transitions. Since the continuous deformation of the contour does not change the value of the integral as long as it does not cross the singularities, one can reduce the integration to that over both sides of the cuts, as shown in~Fig.~\ref{fig:countour}. 

To proceed further it is convenient to consider just a single singularity in the right half-plane of the complex baryon chemical potential. We will do so in the next subsection.

\subsection{Asymptotic behavior of the Fourier coefficients for a function with a branch point }
\label{sec:A}
\begin{figure}
    \centering
    \includegraphics{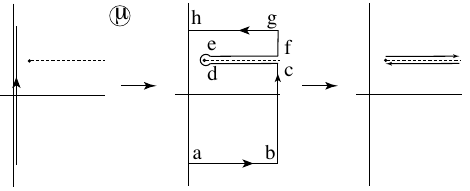}
    \caption{Computation of the Fourier coefficient for the function with one singularity in the right half-plane of complex chemical potential.}
    \label{fig:2}
\end{figure}

Consider an odd  function $n_B (\hat \mu)$ periodic in an imaginary argument having  brunch points in the complex plane located at 
$\pm \hat  \mu^{\rm br}$, where 
$ \hat   \mu^{\rm br} = \hat  \mu_r^{\rm br} + i\, \hat  \mu_i^{\rm br} $.
Here, we assume that near the branch point 
  $\hat \mu \to + \hat  \mu^{\rm br}$:  
$$\hat n_B(\hat \mu)  =  A (\hat \mu - \hat \mu^{\rm br})^{\sigma} 
(1 + B(\hat \mu - \hat \mu^{\rm br})^{\theta_c} + \ldots )  + 
\sum\limits_{n=0}^{\infty} a_n (\hat \mu - \hat \mu^{\rm br})^n
$$ with $\sigma>-1$ and $\theta_c>0$. In the context of the YLE singularity, $\theta_c$ is the confluent critical exponent (not to be confused with $\theta$).   The regular part of the function is encoded in the coefficients $a_n$. 
Our goal is to find the asymptotic behavior of the Fourier coefficients for such a function. 

From the definition of the Fourier coefficients 
\begin{align}
    b_k = \frac{-i}{\pi} \int_{-\pi}^{\pi} d\theta  \,  \hat  n_B (\hat \mu = i \theta ) 
    \sin \left(k \theta \right)
\end{align}
where instead of  ${\rm Im}\,  \hat  n_B (\hat \mu = i \theta ) $ we use a more appropriate   $- i\,  \hat  n_B (\hat \mu = i \theta ) $. Next, using the property $\hat  n_B (\hat \mu) = - \hat n_B (-\hat \mu)$ we further simplify   
\begin{align}
    b_k = \frac{-1}{2\pi} \int_{-\pi}^{\pi} d\theta  \,  \hat  n_B (\hat \mu = i \theta ) 
   \left( e^{i k \theta} - e^{-i k \theta} \right) = \frac{1}{\pi} \int_{-\pi}^{\pi} d\theta  \,  \hat n_B (\hat \mu = i \theta ) 
   e^{-i k \theta} \,.
\end{align}
This is the form convenient for our analysis. To compute the integral, we will deform the contour as shown in Fig.~\ref{fig:2}. In the figure, we assume that the right-most points  are extended to the infinity, i.e. ${\rm Re} \, \mu \to \infty$. The contribution of the segments $(ab)$ and $(gh)$ cancel each other due to the periodicity of the integrand and the opposite direction of the segments. 
The contributions from $(bc)$ and $(fg)$ is zero due to the exponential decay of $\exp(- i k \theta  ) 
= \exp(- k \hat \mu  ) $ for any $k>0$. The integral around the branch point  $(de)$  is vanishing due to $\sigma>-1$. Thus, the only non-trivial contribution is due to the segments on both sides of the cut $(cd)$ and $(ef)$  as demonstrated in the figure. 
To evaluate the contribution of these segments, we consider the parametrization for the segment 
$i \theta = s  + \hat \mu^{\rm br}$
\begin{align}
    \frac{1}{\pi} \int_{(ef)} d\theta  \,  \hat n_B (\hat \mu = i \theta ) 
   e^{-i k \theta} 
   &=     \frac{1}{i \pi} e^{-\mu^{\rm br} k } \int_{0}^{\infty} d s  \,  \hat n_B (\hat \mu = s  + \hat \mu^{\rm br} ) e^{- k s} \,.
\end{align}
Now for an arbitrary power $p>-1$, 
\begin{align}
   \int_{0}^{\infty} d s  \,  s^p  e^{- k s}  = \frac{\Gamma(p+1) }{k^{p+1}}\,. 
\end{align}
Therefore, we get 
\begin{align}
 \nonumber
    \frac{1}{\pi} \int_{(ef)} d\theta  \,  \hat  n_B (\hat \mu = i \theta ) 
   e^{-i k \theta} 
   &=  \frac{ e^{-\mu^{\rm br} k }}{i \pi}  \left(
   A \frac{\Gamma(1+\sigma)}{k^{1+\sigma}} \left[ 
    1 +   \frac{B}{k^{\theta_c}}   \frac{\Gamma(1+\sigma + \theta_c)}{\Gamma(1+\sigma)} 
   + \ldots \right]
    \right. \\ &\left. 
    + \sum_{n=0}^\infty a_n \frac{\Gamma(1+n)}{k^{1+n}} 
   \right)     \,.
\end{align}
In the second line, the contribution is due to the analytic part. 

The integral over the segment $(cd)$ is identical to the expression above except for the $2\pi$ rotation around the branch point and an extra minus sign due to the direction of the segment: 
\begin{align}
\nonumber
    \frac{1}{\pi} \int_{(cd)} d\theta  \, \hat  n_B (\hat \mu = i \theta ) 
   e^{-i k \theta} 
   &=  - \frac{ e^{-\mu^{\rm br} k }}{i \pi}  \left(
   A \frac{\Gamma(1+\sigma)}{k^{1+\sigma}} e^{2\pi \sigma} \left[ 
    1 +   \frac{B}{k^{\theta_c}} e^{2\pi \theta_c}  \frac{\Gamma(1+\sigma + \theta_c)}{\Gamma(1+\sigma)} 
   + \ldots \right]
    \right. \\ &\left. 
    + \sum_{n=0}^\infty a_n \frac{\Gamma(1+n)}{k^{1+n}} 
   \right)     \,.
\end{align}
Adding both integrals together cancels the analytic part to yield 
\begin{align}
    b_k = 
    \frac{ e^{-\mu^{\rm br} k }}{i \pi}  
   A \frac{\Gamma(1+\sigma)}{k^{1+\sigma}} \left(
    1 - e^{i 2\pi\sigma} +   \frac{B}{k^{\theta_c}} \left[ 1 - e^{i 2\pi(\sigma+\theta_c)} \right]   \frac{\Gamma(1+\sigma + \theta_c)}{\Gamma(1+\sigma)} 
   + \ldots \right)
\end{align}
Absorbing $k$ independent factors into constants $A$ and $B$ we finaly have 
\begin{align}
    b_k = 
      \tilde A \frac{e^{-\mu^{\rm br} k }}{k^{1+\sigma}} \left(
    1  +   \frac{\tilde B}{k^{\theta_c}}  
   + \ldots \right)
\end{align}

Now we are ready to generalize our result for two branch points corresponding to RW and chiral YLE singularities. 

\subsection{Asymptotic behavior of the Fourier coefficients in QCD}

Generalizing the result obtained in Sec.~\ref{sec:A} to the case when both YLE and RW singularities are present and accounting for the analytical properties of the partition function (singularities comes in complex conjugate pairs), we obtain   
\begin{align}
    b_k = 
      \tilde A_{\rm YLE} \frac{e^{-\hat \mu^{\rm YLE} k }}{k^{1+\sigma}} \left(
    1  +   \frac{\tilde B_{\rm YLE}}{k^{\theta_c}}  
    + \ldots \right)
    + \tilde A_{\rm RW} \frac{e^{-\hat \mu^{\rm RW} k }}{k^{1+\sigma}} \left(
    1  +   \frac{\tilde B_{\rm RW}}{k^{\theta_c}}  
   + \ldots \right) + {\rm c.c.}\,.
\end{align}
Here the coefficients $\tilde A_{\rm YLE, RW}$ and $\tilde B_{\rm YLE, RW}$ are complex numbers in general. 

Taking into account that ${\rm Im}\, \hat \mu^{\rm RW} =  \pi$, we have 
\begin{align}
\nonumber
    b_k &= 
      |\tilde A_{\rm YLE}| \frac{e^{-\hat \mu_r^{\rm YLE} k }}{k^{1+\sigma}} \left(
    \cos(\hat \mu_i^{\rm YLE} k + \phi^{\rm YLE}_a)  +   \frac{|\tilde B_{\rm YLE}|}{k^{\theta_c}}
    \cos(\hat \mu_i^{\rm YLE} k + \phi^{\rm YLE}_b)
    + \ldots \right) \\ &
    + |\hat A_{\rm RW}|  (-1)^k \frac{e^{-\hat \mu_r^{\rm RW} k }}{k^{1+\sigma}} \left(
    1  +   \frac{|\hat B_{\rm RW}|}{k^{\theta_c}}  
   + \ldots \right)\, ,
\end{align}
where $\phi_a$ and $\phi_b$ are phases due to non trivial phases of $\tilde A^{\rm YLE}$ and $\tilde B^{\rm YLE}$ and trivial real factors were absorbed into $|\hat A_{\rm RW}|$ and $|\hat B_{\rm RW}|$. 
This is the final expression. 
Note that coefficients, $b_k$, are exponentially sensitive to the imaginary values of the positions of the YLE singularities.  

\begin{figure}
    \centering
    \includegraphics[width=0.5\linewidth]{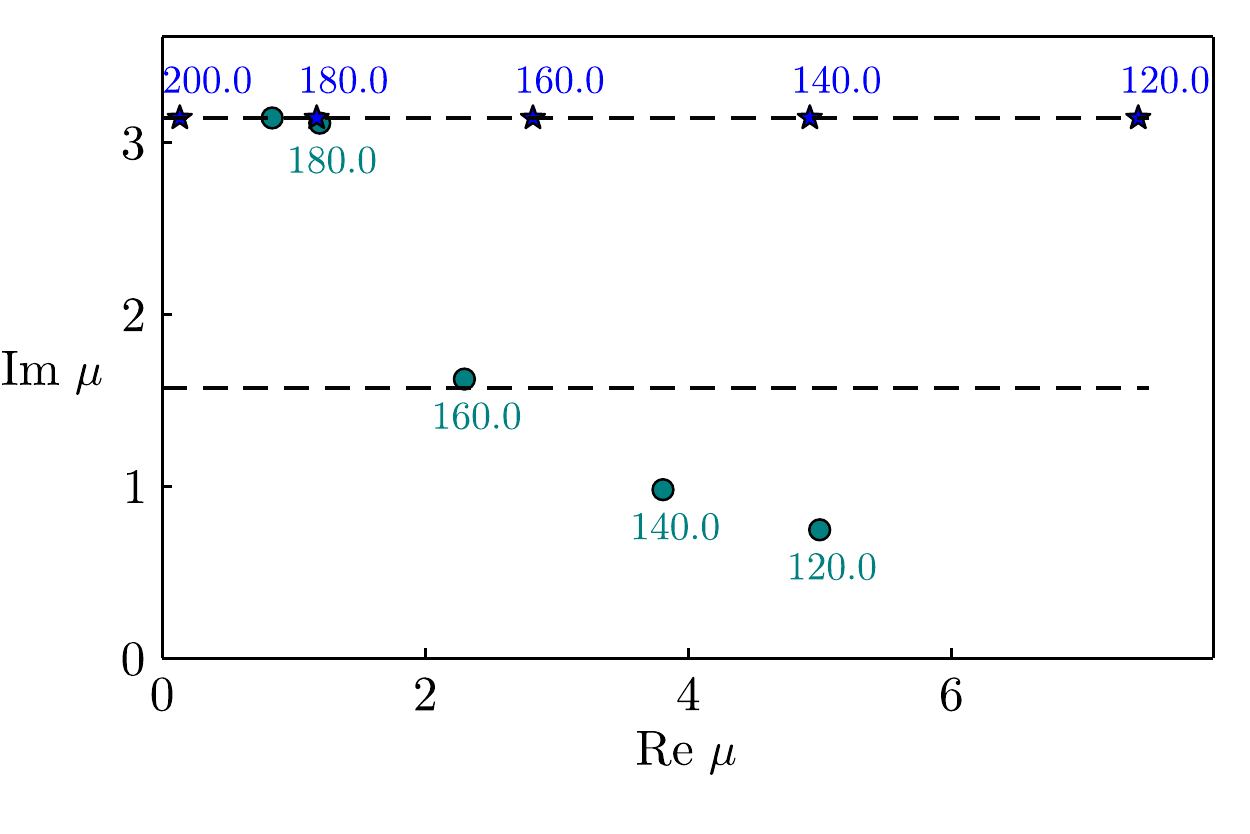}
    \caption{Illustration of the positions of the singularities under the scaling assumptions in QCD. The stars (dots) indicate
    the positions of RW (chiral) YLE. Numerical labels indicate the corresponding temperatures in MeV.  
    The horizontal dashed lines indicate ${\rm Im} \hat \mu = \pi/2$ and $\pi$.} 
    \label{fig:singQCD}
\end{figure}

The confluent critical  exponent $\theta_c = \nu_c \omega = \frac{\sigma+1}{3} \omega$  is about $0.6$~\footnote{Here we used the value from five-loop $\varepsilon$ expansion~\cite{Borinsky:2021jdb}, $\omega \approx 1.6$.} and thus leads to an appreciable suppression of the corrections. It is thus safe to drop them: 
\begin{align}
\label{Eq:fit}
    b_k &= 
      |\tilde A_{\rm YLE}| \frac{e^{-\hat \mu_r^{\rm YLE} k }}{k^{1+\sigma}}
    \cos(\hat \mu_i^{\rm YLE} k + \phi^{\rm YLE}_a) 
    + |\hat A_{\rm RW}|  (-1)^k \frac{e^{-\hat \mu_r^{\rm RW} k }}{k^{1+\sigma}}
\end{align}

\begin{figure}
\begin{subfigure}{0.45\textwidth}
    \centering
  \includegraphics[width=0.95\linewidth]{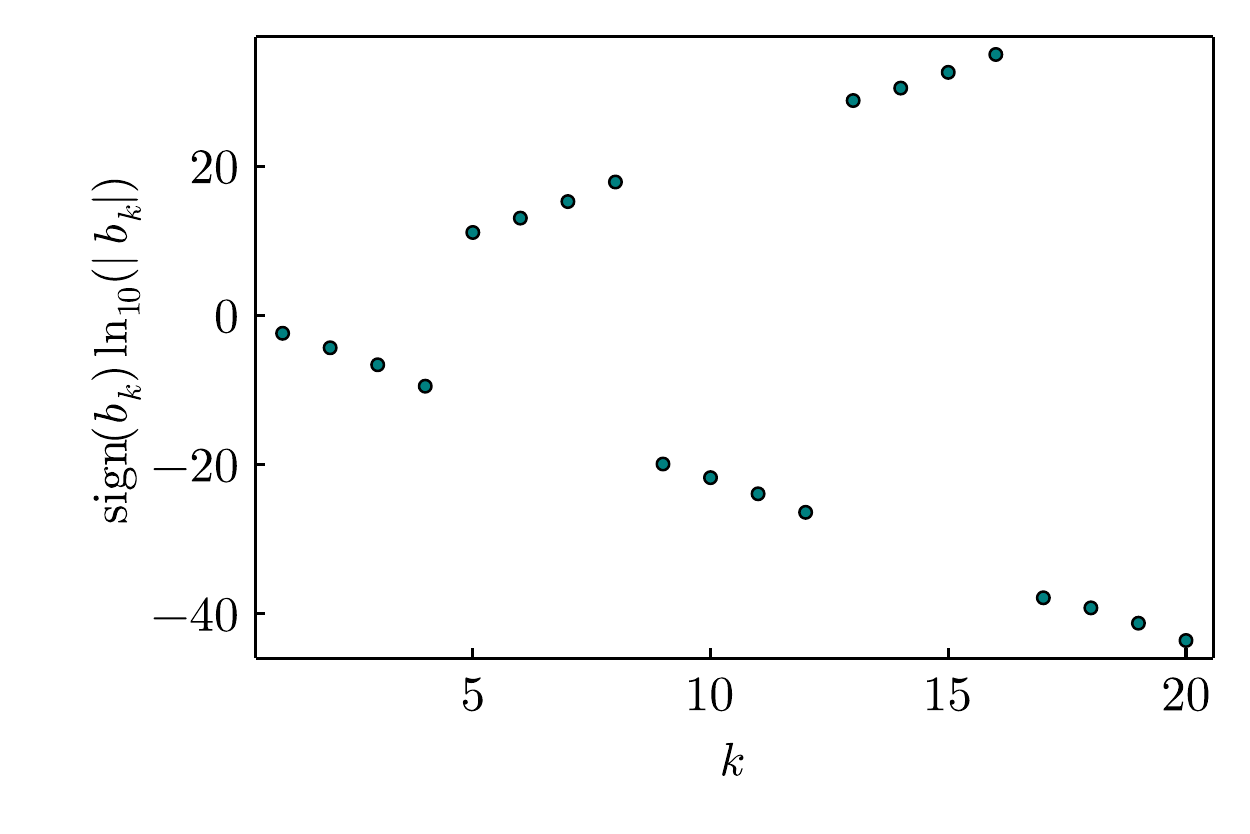}
    \caption{$T=120$  MeV}
    \label{fig:coeffQCD1}
\end{subfigure}
\begin{subfigure}{0.45\textwidth}
    \centering
    \includegraphics[width=0.95\linewidth]{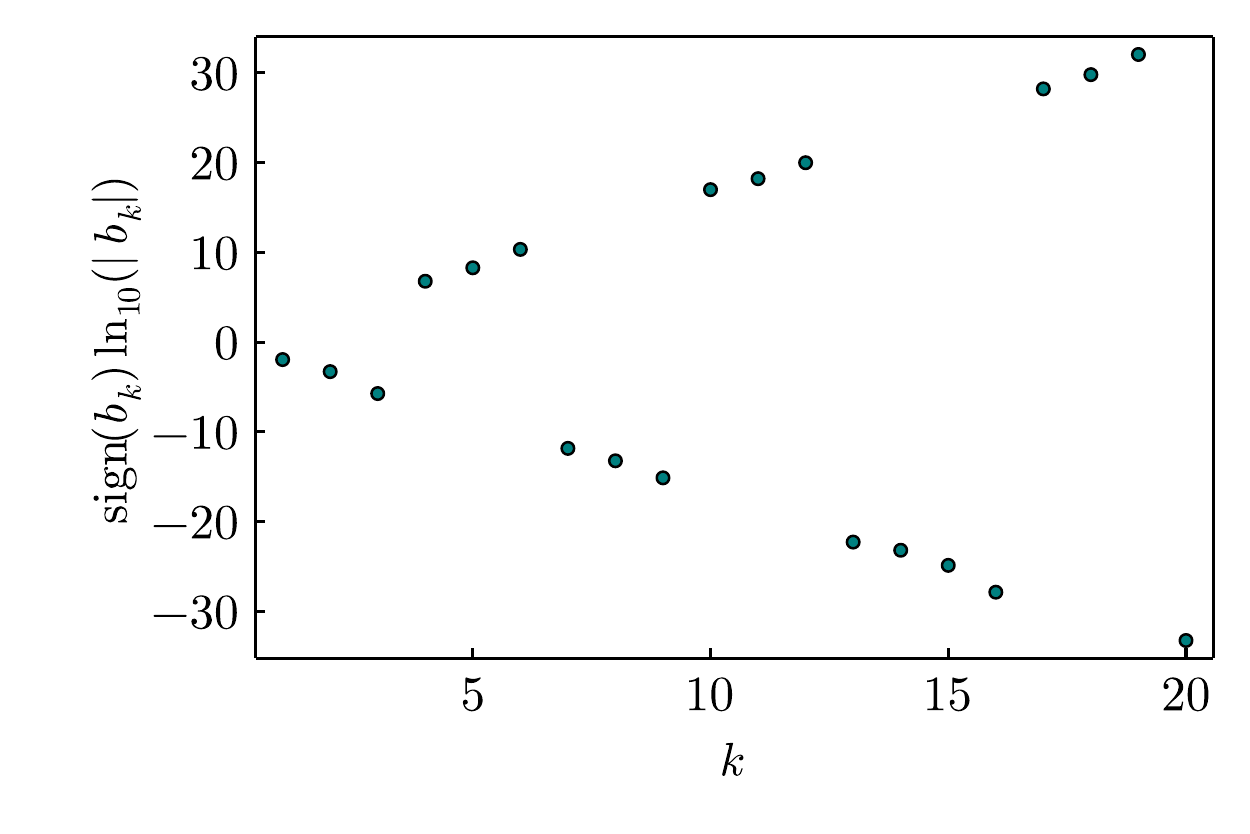}
    \caption{$T=140$ MeV} 
    \label{fig:coeffQCD2}
\end{subfigure}
\begin{subfigure}{0.45\textwidth}
    \centering
  \includegraphics[width=0.95\linewidth]{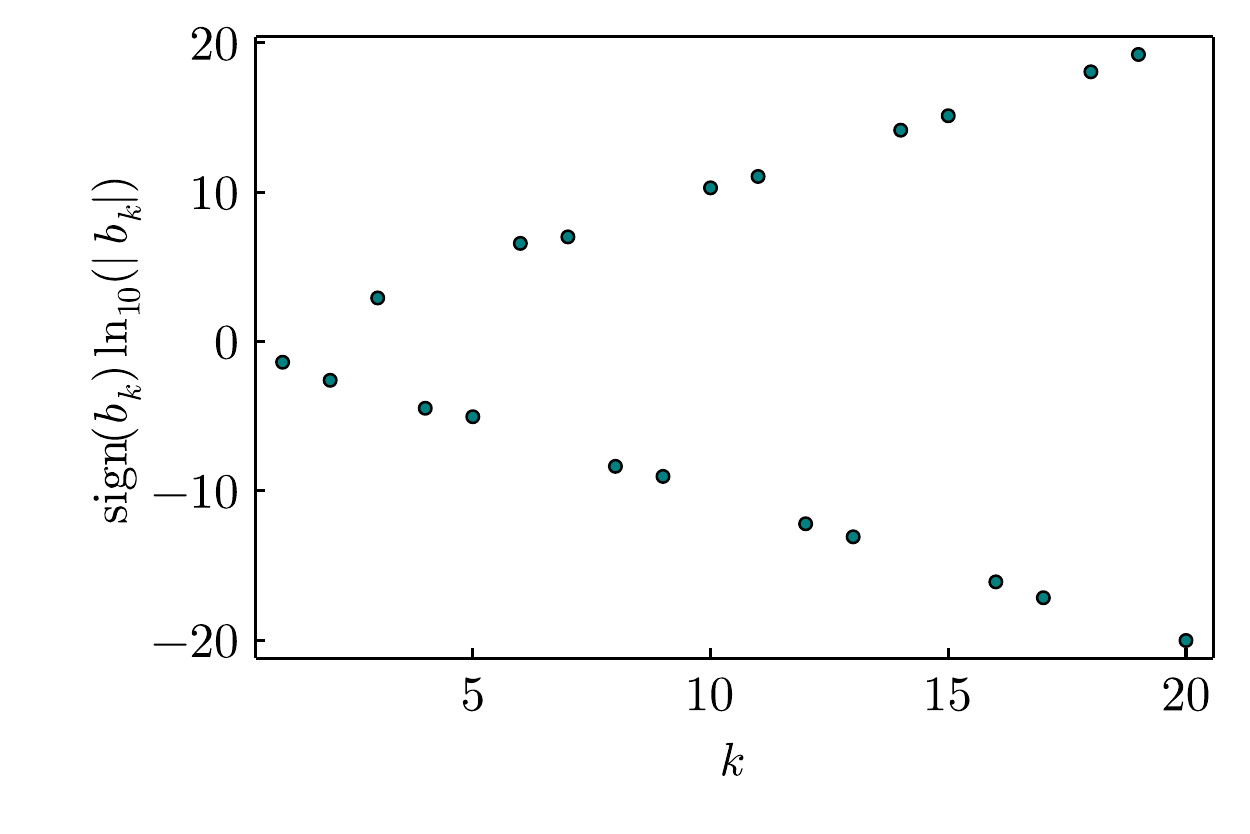}
    \caption{$T=160$  MeV}
    \label{fig:coeffQCD3}
\end{subfigure}
\begin{subfigure}{0.45\textwidth}
    \centering
    \includegraphics[width=0.95\linewidth]{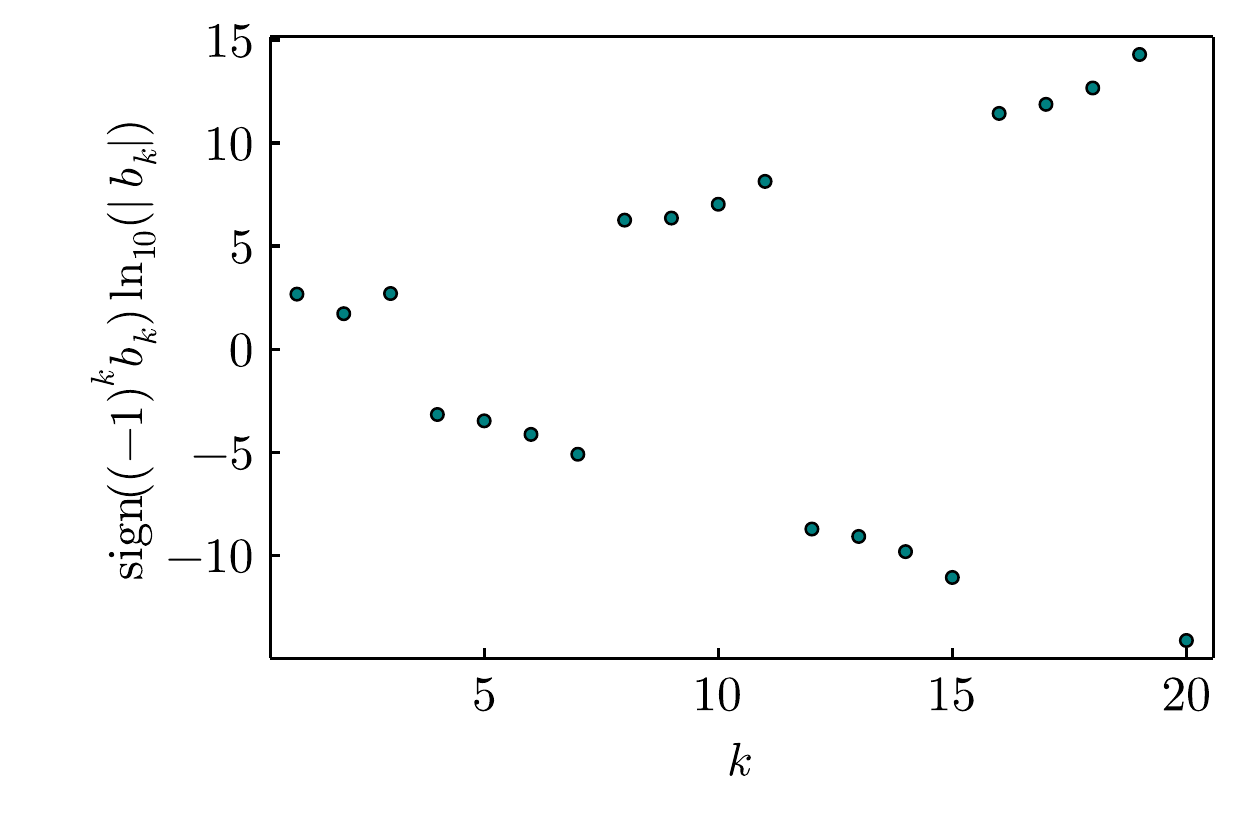}
    \caption{$T=170$ MeV} 
    \label{fig:coeffQCD4}
\end{subfigure}
\begin{subfigure}{0.45\textwidth}
    \centering
  \includegraphics[width=0.95\linewidth]{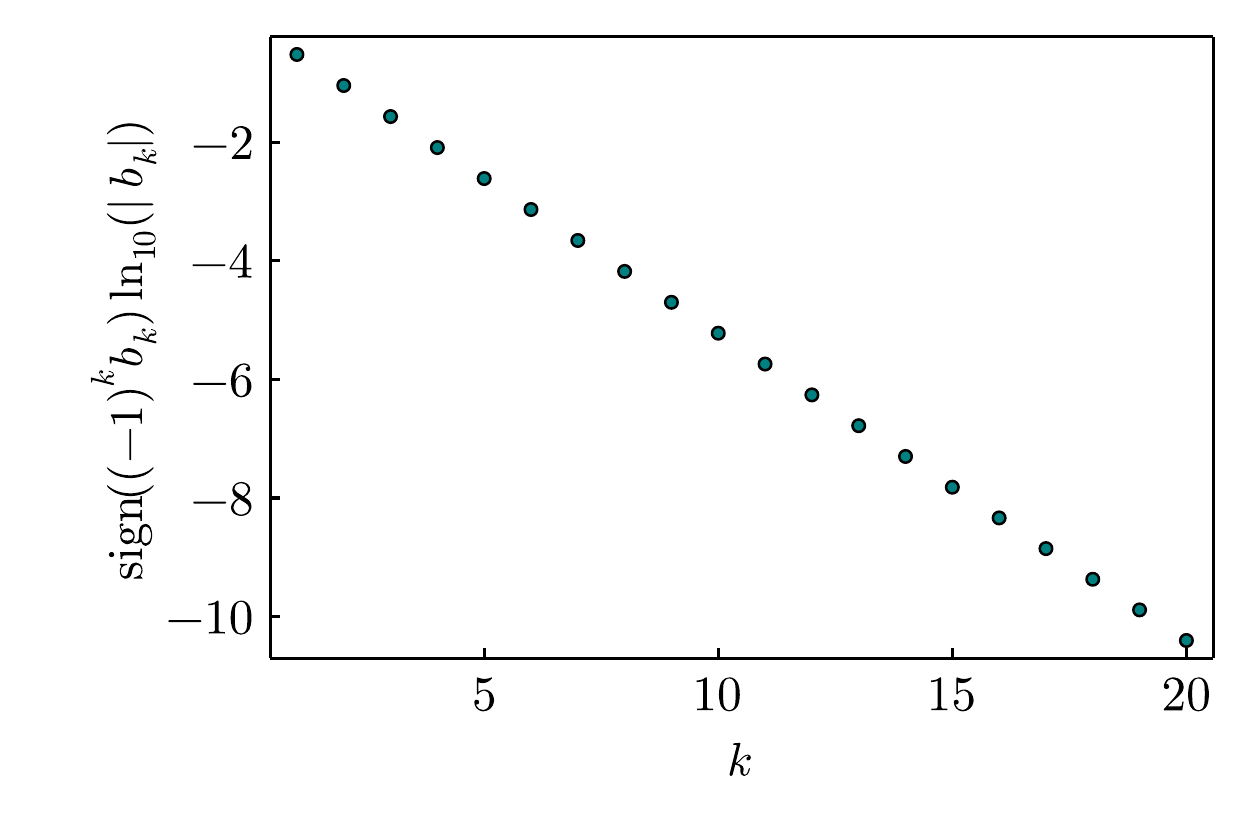}
    \caption{$T=180$  MeV}
    \label{fig:coeffQCD5}
\end{subfigure}
\begin{subfigure}{0.45\textwidth}
    \centering
    \includegraphics[width=0.95\linewidth]{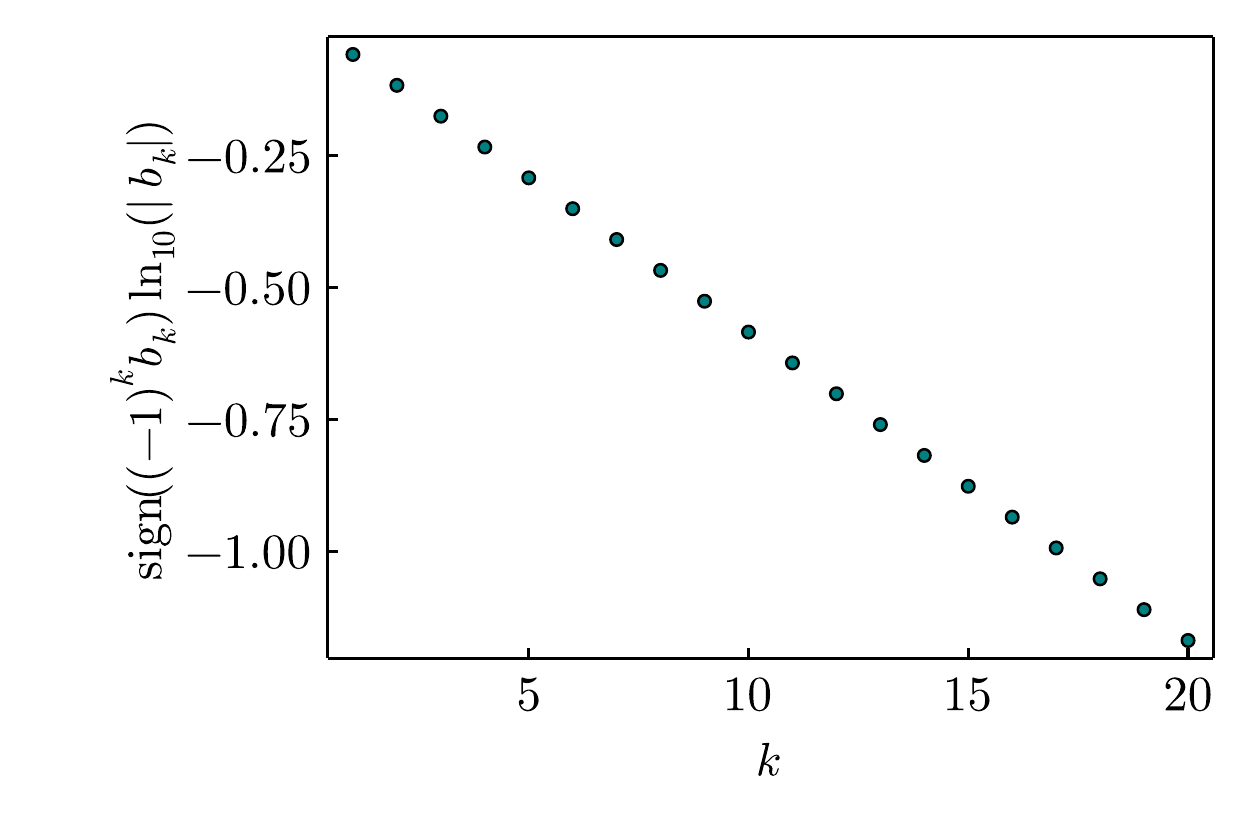}
    \caption{$T=200$ MeV} 
    \label{fig:coeffQCD6}
\end{subfigure}
        \caption{Illustration of different qualitative behaviour of the Fourier coefficients. Note that although we plotted the coefficients at small values of $k$, Eq.~\eqref{Eq:fit} is only formally valid for asymptotically large $k$. }
        \label{Fig:Coeff}
\end{figure}

When the chiral YLE singularity approaches ${\rm Im } \, \hat \mu = \pi/2$, the sign of $b_k$ starts to change very rapidly. Indeed 
for a special case of  $\hat \mu_i^{\rm YLE} = \pi/2$ and zero phase,  
one get $\cos \frac{\pi k}{2}$, which is zero for any odd $k$, positive for  $k=0,4,8,..$ and negative $k=2,6,10$.   
In this case it is convenient to multiply $b_k$ by $(-1)^k$ to get 
\begin{align}
  (-1)^k k^{1+\sigma}  b_k &= 
      |\tilde A_{\rm YLE}| e^{-\hat \mu_r^{\rm YLE} k }
    \cos( (\pi - \hat \mu_i^{\rm YLE}) k - \phi^{\rm YLE}_a) 
    + |\hat A_{\rm RW}| e^{-\hat \mu_r^{\rm RW} k }\,.
\end{align}
This shifts the frequency of the oscillations to lower values. 

To illustrate the behavior of the Fourier coefficients, we consider the scaling assumption for the position of the Yang-Lee edge singularities. For that we will use the input from Refs.~\cite{Connelly:2020gwa,Rennecke:2022ohx,Johnson:2022cqv} on the universal locations, well known critical exponents for Z(2) and O(4) universality classes and approximation for the scaling variables
$z = z_0\, t/h^{1/\beta \delta}$ with 
\begin{align}
t &= \frac{T_{\rm RW}-T}{T_{\rm RW}}, \quad h = \frac{\mu_B-i\pi}{i\pi} \quad \text{for RW}, \\  
t &= \frac{T-T_c}{T_{c}} + \kappa^B_{2} \left(\frac{\mu}{T} \right)^2, \quad h = \frac{m_{u,d}}{m_{s}} \quad \text{for chiral phase transition}\,.
\end{align} 
The non-universal coefficients  ($T_{\rm RW}$, $T_{c}$, $\kappa_2^B$, and $z_0$ for both transitions) are taken from lattice QCD parametrizations, see Refs.~\cite{Clarke:2023noy}.  

Using this asymptotic formula and the scaling assumption for the locations of the singularities (the position is displayed in Fig.~\ref{fig:singQCD}) we plotted the qualitative behavior of the Fourier coefficients in Fig.~\ref{Fig:Coeff}, see also a detailed description of the approach in application to the chiral phase transition in Ref.~\cite{Mukherjee:2019eou}. 

Lets examine the figures. Consider Fig.~\ref{fig:coeffQCD1}, the period of the oscillations (sign changes in $k$) is about $P_k = 4$; we thus can easily estimate the value of $\hat \mu_i^{\rm YLE}$ to be  $\pi/P_{k}=\pi/4$. This estimate is close the input value of  $\hat \mu_i^{\rm YLE} \approx  0.24\, \pi$. Similarly for Fig.~\ref{fig:coeffQCD2}, we get an estimate of $\pi/3$ to be compared with the input value of $\hat \mu_i^{\rm YLE} \approx  0.312\, \pi$. In Fig.~\ref{fig:coeffQCD4}, we multiplied the Fourier coefficient by $(-1)^k$, as we explained before it helps to lower the frequency of the sign change. In this case we need to adjust our estimate, as the period of the sign change will be related to the imaginary value of the YLE chemical potential trough  $\pi - \hat \mu_i^{\rm YLE} = \pi /P_k$, we thus have an estimate $3\pi/4$ for  $\hat \mu_i^{\rm YLE}$. This is to be compared to the input value of $0.74\, \pi$. For higher temperatures, the sign change is absent in $(-1)^k b_k$ demonstrating the fact that the singularities are on the line  $\hat \mu_i = \pi$. 

The real value of the location of the YLE singularity (at smaller $T$) and RW singularity (at larger $T$) can be read off from the slopes in the figures.  We thus see that a simple analysis of the Fourier coefficients provide a good estimate for the location of the singularities.


%

\section{Model results}
\label{Sec:MR}

In this section we perform the calculations of the Fourier coefficient in Quark-meson model 
in both mean-field approximation  and by accounting for fluctuations in the local potential approximation of the Functional Renormalization Group approach (for the later, the critical exponent at YLE singularity is numerically closer to the actual one, see Introduction). We will demonstrate that the analysis of a limited number of Fourier coefficients complemented by the asymptotic expression derived in the previous subsection is sufficient to extract the location of the Yang-Lee edge singularity with a good precision. 

\subsection{Mean-field quark-meson model}
\label{sec:MF}

In this section, we follow the notation of Ref.~\cite{Mukherjee:2021tyg}.
The Euclidean action  for a quark-meson model with $N_f=2$ degenerate quark flavors and  $N_c = 3$ colors is given by 
\begin{align}\label{eq:lag}
\begin{split}
S^{\rm QM} &= \int_0^\beta\! dx_0 \int\! d^3x\,\bigg\{ \bar\psi \bigg( \gamma_\mu\partial_\mu+ \frac{1}{2} h\, \bm{\tau} \bm{\phi} + \gamma_0 \mu \bigg)\psi  + \frac{1}{2} (\partial_\mu \bm\phi)^2 + U(\bm{\phi}^2) - h \sigma\bigg\}\,.
\end{split}
\end{align}
$\gamma_\mu$ are the Euclidean gamma matrices, $\bm{\tau}^T = (1,i\gamma_5 \vec{\tau})$ with the Pauli matrices $\vec{\tau}$, $\mu$ is the quark chemical potential, and $\bm{\phi}^T = (\sigma,\vec{\pi})$ is the $O(4)$ meson field. $U(\bm{\phi}^2)$ is the effective meson potential. An explicit symmetry breaking is introduced through the source $h$ to get massive pions in the low-temperature phase. 

Assuming a homogeneous mean field with the non-zero expectation value for the iso-singlet meson ($\bar \sigma$) we obtain the following  thermodynamic potential 
\begin{align}
\begin{split}
\bar\Omega^{\rm QM}(T,\mu;\bar\sigma) 
=U(\bar{\sigma}^2) - h\bar\sigma - \frac{T}{V} \ln\det M^{\rm QM}(\bar\sigma;\mu)\,,
\end{split}
\end{align}
where 
\begin{align}
\begin{split}
 &\frac{T}{V} \ln\det M^{\rm QM}(\bar\sigma;\mu) =  2 N_f N_c \Big[ J_0(\bar\sigma) + J_{T,\mu}(\bar\sigma) + J_{T,-\mu}(\bar\sigma)  \Big]\,.
\end{split}
\end{align}
Here  $\int_q = \int\!\frac{d^3q}{(2\pi)^3}$ and we defined the thermal contribution to the quark determinant as
\begin{align}\label{eq:JT}
J_{T,\mu}(\bar\sigma) = \frac{1}{2 \pi^2} \int_0^\infty\!dq\, q^2\, T\, \ln\Big[ 1 + e^{-(E_q(\bar\sigma)-\mu)/T} \Big]\,.
\end{align}
The vacuum contribution $J_0$ is ultraviolet-divergent. Its finite piece  depends on the meson field (see Ref.~\cite{Skokov:2010sf})
\begin{align}\label{eq:J0eps}
J_0^\epsilon(\bar\sigma) = \frac{h^4\bar\sigma^4}{2^9 \pi^2} \ln\bigg(\frac{h^2\bar\sigma^2}{4 \Lambda^2}\bigg)\,.
\end{align}
For the O(4)-symmetric part of the meson potential we use 
\begin{align}
U(\bm{\phi}^2) = \frac{\lambda}{4} \big(\bm{\phi}^2-\nu^2\big)^2\,,
\end{align}
which allows for spontaneous symmetry breaking. 
Following the logic of the mean-field approximation, physical results are extracted the minimum of the thermodynamic potential, that is 
\begin{align}\label{eq:barOmegaQM2}
\bar\Omega^{\rm QM}(T,\mu)=\bar\Omega^{\rm QM}(T,\mu;\bar\sigma_0)\,,
\end{align}
where $\bar\sigma_0$ is the solution of the equation of motion
\begin{align}\label{eq:QMEoM}
\frac{\partial \bar\Omega^{\rm QM}(T,\mu;\bar\sigma)}{\partial \bar\sigma}\bigg|_{\bar\sigma_0} = 0\,.
\end{align}

Further details of the model can be found in Ref.~\cite{Skokov:2010sf}. We used the following input values to fix the parameters of the model $m_\sigma = 600$ MeV, $m_\pi = 140$ MeV, $f_\pi = 93$ MeV and the Yukawa coupling is fixed to be $g=3.6$.   

\begin{figure}
    \centering
    \includegraphics[width=0.49\linewidth]{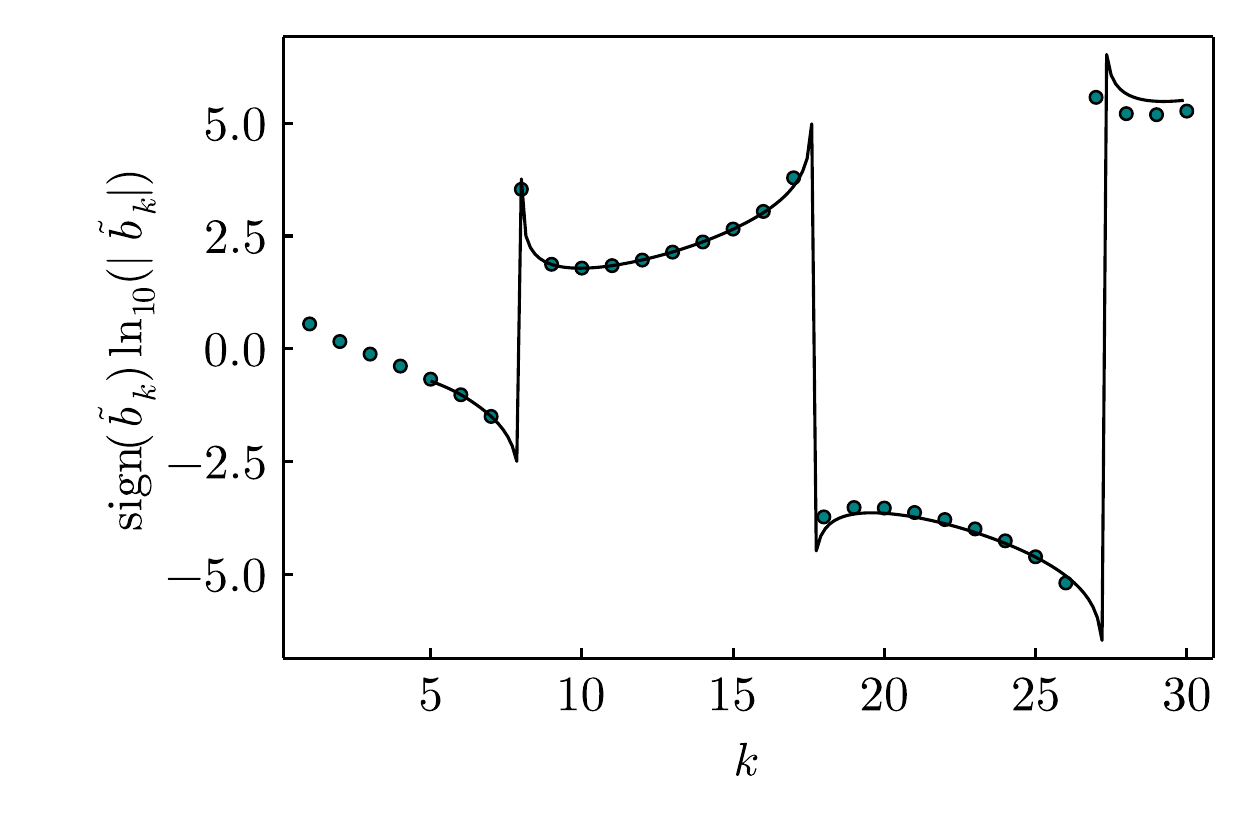}
    \includegraphics[width=0.49\linewidth]{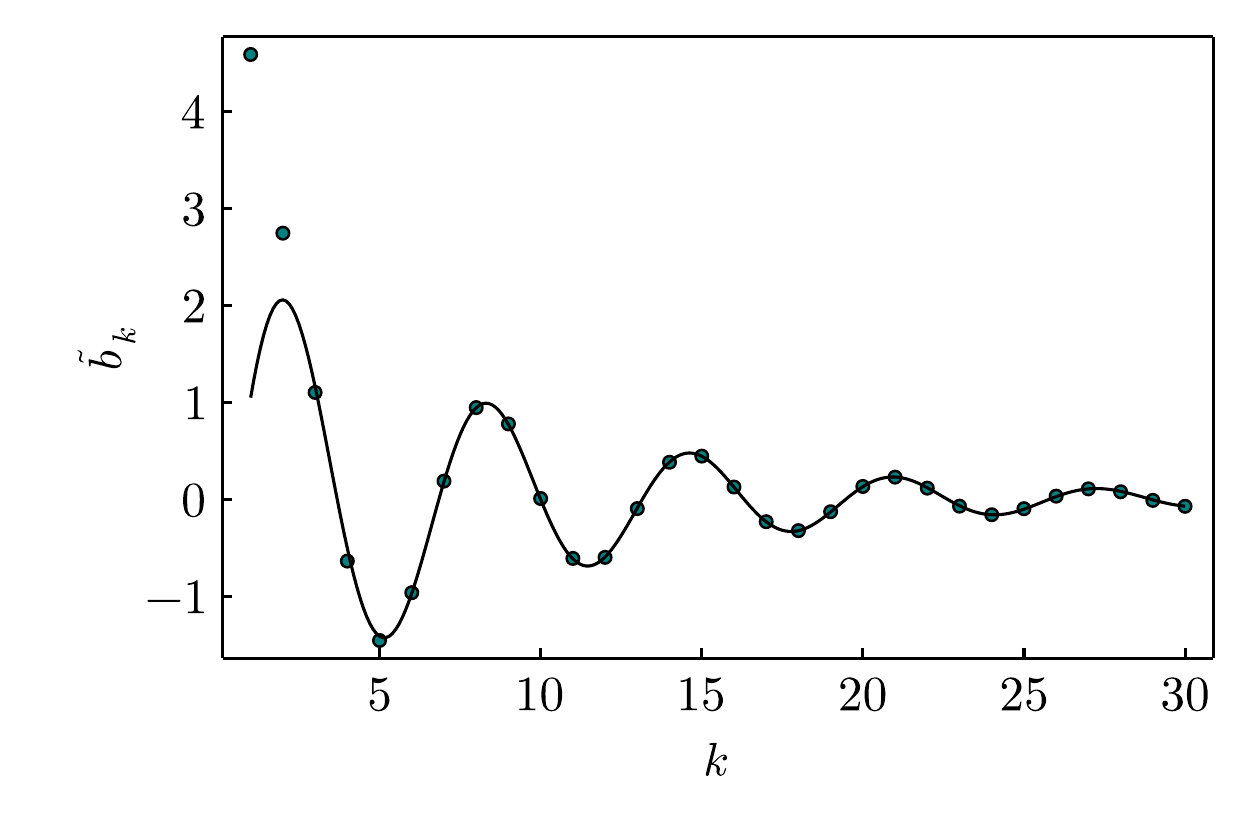}
    \caption{Mean-field Fourier coefficients $\tilde b_k = k^{1+\sigma} b_k$ ($\sigma_{\rm MF} = 1/2$) for $T = 150$ MeV (left) and $T = 180$ MeV (right)   and the corresponding fits. 
}
    \label{fig:mf}
\end{figure}

Figure~\ref{fig:mf} summarizes our results. The data was fitted using~Eq.~\eqref{Eq:fit}. The results of the fit are as follows.  The fit yields $\hat \mu^{\rm fit}_{\rm YLE} = 0.441(2) + i\, 0.325(3)$ 
($\hat \mu^{\rm fit}_{\rm YLE} = 0.1156(6) + i\, 0.9952(5)$)
for $T = 150 (180)$  MeV.      The actual location of the singularity is $\hat \mu_{\rm YLE} = 0.412884 + i\, 0.342187$ ($\hat \mu_{\rm YLE} = 0.118657 + i\, 1.00256$)  for $T = 150 (180)$  MeV.
The fit reproduces the location within 7\% precision. 
The fits are performed starting from $b_5$ and do not require going to asymptotically large Fourier modes. Note that the imaginary part of the chemical potential can be estimated easily without performing the fits and visually extracting the period of the sign change; for example, for $T=180$ MeV, the sign changes every three points, this leads to $\hat \mu_{\rm YLE} \approx \pi/3  \approx 1.05$, which is very close to the actual value. 

\subsection{Quark-meson model in LPA FRG }
The ingredients of the model are similar to those described in the previous subsection. The model is computed using FRG in the local potential approximation. This approximation leads to $\sigma_{\rm YLE} = 1/5$, which is closer to the actual value. 
The model is described in detail in Ref.~\cite{Skokov:2010wb}. We used the same values for $m_\sigma$, $m_\pi$,  $f_\pi$,  and $g$ as in the mean-field model. Additionally, we fixed the UV cut-off (a new ingredient required by the FRG approach) to 1200 MeV.  
 
\begin{figure}
    \centering
    \includegraphics[width=0.49\linewidth]{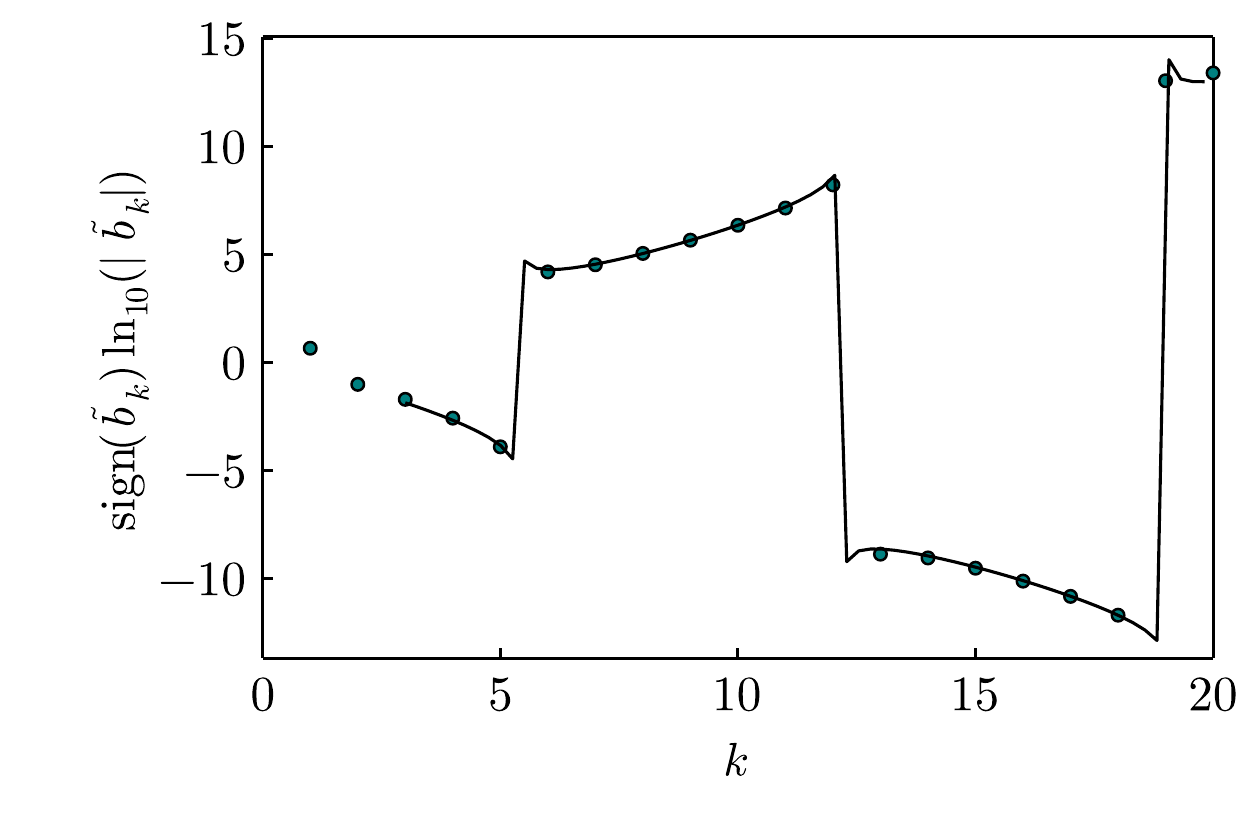}
    \includegraphics[width=0.49\linewidth]{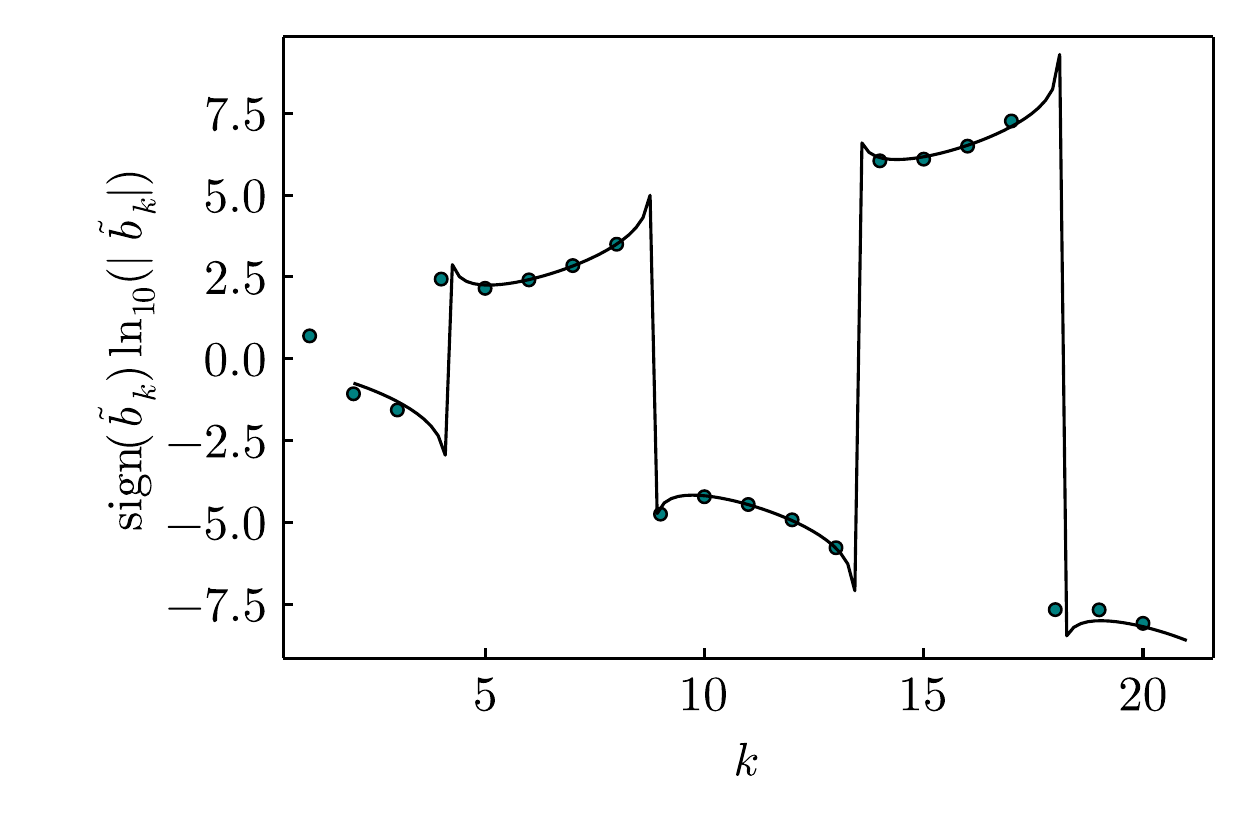}
    \caption{LPA FRG  Fourier coefficients $\tilde b_k = k^{1+\sigma} b_k$ ($\sigma_{\rm LPA} = 1/5$)  and the corresponding fits for $T = 150$ MeV (left) and $T = 180$ MeV (right).
    }
    \label{fig:frg}
\end{figure}

Figure~\ref{fig:frg} summarizes our result. The fit yields $\hat \mu^{\rm fit}_{\rm YLE} = 1.483(7) + i\, 0.446(6)$ 
($\hat \mu^{\rm fit}_{\rm YLE} = 0.949(8) + i\, 0.675(11)$)
for $T = 150 (180)$  MeV.  The actual location of the singularity is $\hat \mu_{\rm YLE} = 1.553 + i\, 0.4794$ ($\hat \mu_{\rm YLE} =  
0.9445 + i\,  0.6618$)  for $T = 150 (180)$  MeV. 
Although we could not reliably extract the Fourier coefficient for $k>22$ (and thus are limited to relatively modest values of $k$), the fit accuracy is sufficiently high as they reproduce the locations within 5\% precision.   
As for the mean field model, the fits start from $b_5$. Again, a straightforward visual inspection of the plot allows us to establish the period of the sign change and thus estimate the location for the YLE singularity. Consider $T=150$ MeV; the period is between 6 and 7. Let's approximate it by 6.5. Thus 
$\hat \mu_{\rm YLE} \approx \pi/ 6.5 \approx 0.48$; this is amusingly close the the actual value.

\section{Conclusions}
\label{Sec:Con}

In this manuscript, we derived the asymptotic behavior of the Fourier coefficients in the crossover regime. Our result differs from the one obtained in Ref.~\cite{Almasi:2019bvl}, where the authors used the Riemann-Lebesgue lemma and where, we believe, the critical exponents and the amplitudes at YLE singularities were misidentified. In this manuscript, we analyzed the behavior of the coefficients and concluded that the position of the YLE (closest to the real axis) singularity can be extracted knowing about 20 Fourier coefficients. Applying our analysis to lattice QCD has two caveats. First, finite size effects alter the asymptotic behavior of the coefficients. This is straightforward to account for, as we demonstrated in Appendix~\ref{sec:AppA}. Second, lattice QCD calculations come with unavoidable statistical errors. They might significantly alter the higher-order Fourier coefficients. In Appendix~\ref{Sec:AppB}, we estimated the upper bound on the error to extract the required coefficient order. A dedicated effort might make achieving this level of accuracy feasible. In Appendix~\ref{Sec:AppC} we propose a specific numerical method, which seems most appropriate for our purpose.      

\acknowledgments

This work 
is supported by the U.S. Department of Energy, Office 
of Nuclear Physics through contract DE-SC0020081 (V.S.).
This work is further supported by the
Deutsche Forschungsgemeinschaft (DFG, German Research Foundation) - Project number 315477589-TRR
211 and the PUNCH4NFDI consortium, supported by
the Deutsche Forschungsgemeinschaft (DFG, German
Research Foundation) with project number 460248186
(PUNCH4NFDI) (C.S.).

V.S. thanks the ExtreMe Matter Institute for partial support and A. Andronic for hospitality at Physics Department of Muenster University, where this project was initiated.  
We (C.S. and V.S) also thank the organizers of the EuroPlex final conference for their support and hospitality. 

\appendix

\section{Finite volume effects}
\label{sec:AppA}
In a finite volume, the singularity and the branch cut are replaced by a countable number of zeros. The first zero approximates the location of the branch cut singularity in the infinite volume limit (the distance from the first zero to the singularity is proportional to $V^{-(1+\sigma)^{-1}}$ \cite{Itzykson:1983gb}). Going to the thermodynamic limit, the density of zeroes close to the edge can either diverge or tend to zero, subject to the sign of the critical exponent $\sigma$. The behavior of the density  can be approximated by 
$g(\mu) \sim (\mu-\mu_{\rm YLE})^\sigma$ since $\sigma>0$ for three-dimensional systems, the density of zeroes approaches zero close to the edge. For the Fourier coefficients, this means that the closest zero (the zero with the smallest absolute value of the real part) will be the dominant contributor, with the corrections suppressed exponentially. This is especially significant for smaller volumes as the distance between the first and the second zero scales as $\sim V^{-\sigma/(1+\sigma)}$. This compels us to consider a finite volume system. For the baryon number density, we will have simple poles in the complex $\mu$-plane located at $\mu_l$, $l=1,2,...$. Performing similar math as in Section~\ref{sec:A} and completing it with complex conjugate zeroes, we get 
\begin{align}
\label{Eq:FV}
    b_k = 
\sum_l   A_l e^{- {\rm Re} (\mu_l) k } \cos({\rm Im} (\mu_l) k + \phi_l) \approx    A_1 e^{- {\rm Re} (\mu_1) k } \cos({\rm Im} (\mu_1) k + \phi_1)
\end{align}
where in the last part, we neglected distant zeroes. This shows that for small volumes, the power-low part of the decay of the Fourier coefficients is gone, and only exponential decay remains. Note that using the behavior of the density of zeroes,  the equality of Eq.~\eqref{Eq:FV} reduces to its infinite volume limit.

\section{Modelling statistical errors of QCD Monte-Carlo simulations}
\label{Sec:AppB}  
In the main body of the paper, we assumed that the Fourier coefficients can be determined with high precision. 
Computing Fourier coefficients using lattice QCD methods leads to unavoidable statistical error (needless to mention systematic errors). 
Here, we will estimate the naive requirement in accuracy for the input data (baryon number), when we attempt to access higher-order Fourier coefficients by means of the standard \textit{Discrete Fourier Transformation}(DFT).  To model statistical error, we will assign uncorrelated Gaussian error to each value of the baryon number density $\hat n_B(\theta_n) \to \hat n_B(\theta_n) +\xi_n$, where $\xi_n$ have the same variance of $\sigma_{\xi}$. We want to determine the upper limit on this quantity, i.e. where the relative error becomes as large as 100\%. 

Here, we consider a uniformly distributed $\theta_n$ range from 0 to $\pi$, as it is required in the standard DFT. Thus, for the Fourier coefficients, we will obtain 
\begin{align}
    b_k^{\rm w/ error} \approx \sum_n  (\hat n_B(\theta_n) +\xi_n) \sin(\theta_n k) =  b_k + \chi_k \, ,
\end{align}
where $\chi_k$ is a Gaussian random number with the variance 
given by $\sigma_\chi^2 = \sigma_{\xi}^2 \sum_n \sin^2(\theta_n k)$. The upper limit on $\sigma_{\xi}$ is then 
\begin{align}
    \sigma_{\xi} < \frac{| b_k|}{\left(\sum_n \sin^2(\theta_n k)\right)^{1/2}} \,.
\end{align}
For any integer $k$ and uniformly samples $\theta_n$, the sum in the denominator can be easily analytically computed and is $k$ independent  $\sum_{n=0}^N \sin^2(\theta_n k) = \frac{N}{2}$, where $N$ is the number of sampled point in $\theta$. We thus obtain that 
\begin{align}
    \sigma_{\xi} < \frac{\sqrt{2} |b_k|}{\sqrt{N}} \,.
\end{align}
To extract $b_{k}$, one has to have $N=k$ at least. Thus, we can get a better estimate
\begin{align}
    \sigma_{\xi} < \frac{\sqrt{2} |b_k|}{\sqrt{k}} \,.
\end{align}
Thus, to extract, say $b_{10}$, assuming that $b_{10}\sim10^{-3}$, the uncertainty in computing the baryon number divided by $T^3$ has to be of order $10^{-3}-10^{-4}$. We note that there exist much more appropriate methods to compute the Fourier coefficients, as we emphasize in Appendix~\ref{Sec:AppC}.

\section{On the numerical calculation of the Fourier Coefficients}
\label{Sec:AppC}  
During the last decade, much progress was made in numerical techniques for highly oscillatory integrals \cite{doi:10.1137/1.9781611975123}. While the standard method for the numerical calculation of the coefficients $b_k$, defined in  Eq.~\eqref{Eq:bk}, is still the \textit{Discrete Fourier Transformation} (DFT), it has two severe limitations: the numerical error grows with the index $k$ of the Fourier coefficient and the number of accessible coefficients is limited by the number of support points. It is thus tempting to investigate one of the newer methods which are asymptotically correct, i.e. by construction the numerical error decreases with $k$. In particular, the piece-wise Filon-type quadrature is making use of $N+1$ function values 
$\hat{n}_{B,j}^{(0)}\equiv\hat{n}_B(i\theta_j)$ at support points $\{i\theta_j:j=0,\dots,N\}$, as well their first few derivatives $\left. \hat{n}_{B,j}^{(s)}\equiv\partial^s \hat{n}_B(\hat\mu_B)/\partial \hat\mu_B^s\right|_{\hat\mu_B=i\theta_j}$, for $s=1,\dots ,S$. The philosophy is to perform a polynomial (Hermite) interpolation on each sub-interval $I_j=[i\theta_j,i\theta_{j+s}]$. The quadrature weights can than be calculated exactly and, most importantly, one can show that the methods becomes asymptotically exact, in the sense that the asymptotic expansion of the error in $1/k$ starts at order $\mathcal{O}(k^{-S-2})$.
This method is directly applicable on the lattice QCD data that is available from simulations at imaginary chemical potentials \cite{Dimopoulos:2021vrk}. However, we note that the error is only decaying polynomially in $1/k$, while the absolute value of the Fourier coefficients are expected to decay exponentially. I.e., the relative error will nevertheless start to grow above a certain order. 

\begin{figure}
    \centering
    \includegraphics[width=0.49\textwidth]{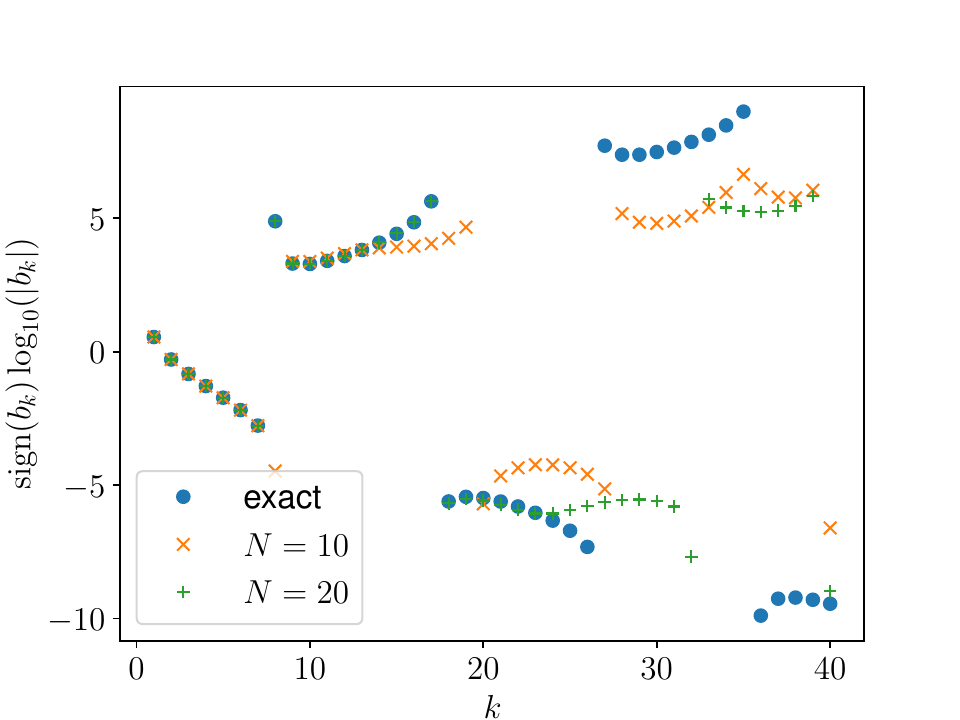}
    \includegraphics[width=0.49\textwidth]{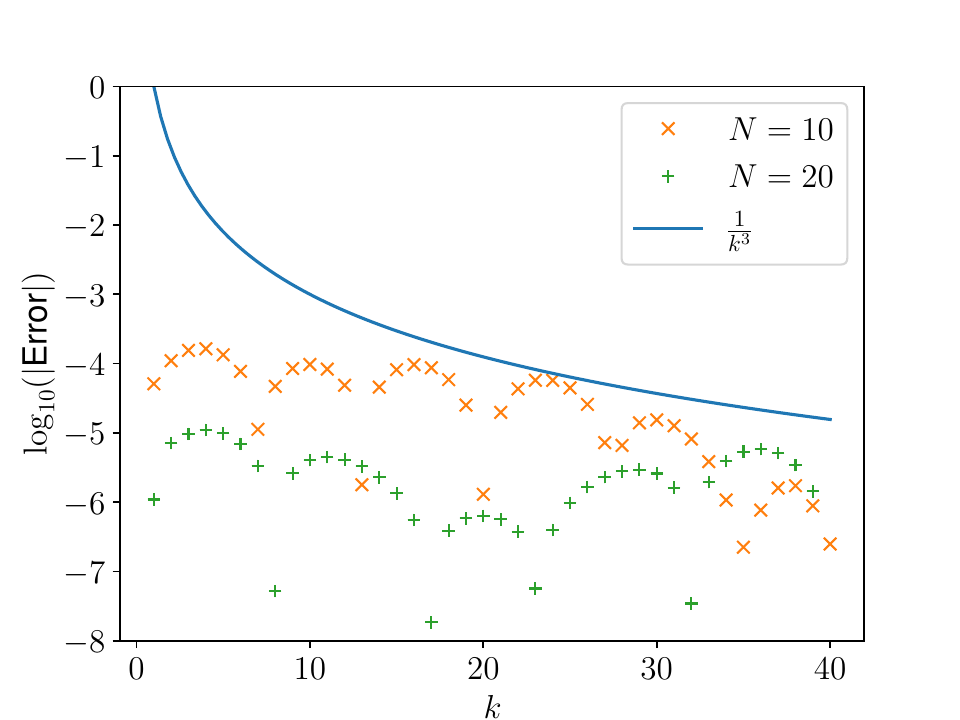}
    \caption{Comparison of the Fourier coefficients from the plain piece-wise Filon quadrature on $N$ sub-intervals with the exact results for the case of the mean-field quark-meson model at $T=150$ MeV. }
    \label{fig:Filon}
\end{figure}

Here we test the plain piece-wise Filon quadrature for $N=10,20$ and $S=1$ for the case of the mean-field quark-meson model at $T=150$~MeV, as discussed in Sec.~\ref{sec:MF}. In Fig.~\ref{fig:Filon} (left) we compare the results with the exact Fourier coefficients, and in Fig.~\ref{fig:Filon} (right) we show the decadic logarithm of the absolute value of the error $\log_{10}|b_k^{\text{exact}}-b_k^{\text{Filon}}|$. The expected asymptotic behaviour of the quadrature is indicated by the solid line. Despite the fact that the absolute value of the error decays as expected, we find that the period of the Fourier coefficients is obscured already at $\mathcal{O}(N)$.
We note that more advanced extended Filon-type quadrature methods do exist, which we leave for future investigations. 

\bibliography{bib}

\begin{thebibliography}{31}%
\makeatletter
\providecommand \@ifxundefined [1]{%
 \@ifx{#1\undefined}
}%
\providecommand \@ifnum [1]{%
 \ifnum #1\expandafter \@firstoftwo
 \else \expandafter \@secondoftwo
 \fi
}%
\providecommand \@ifx [1]{%
 \ifx #1\expandafter \@firstoftwo
 \else \expandafter \@secondoftwo
 \fi
}%
\providecommand \natexlab [1]{#1}%
\providecommand \enquote  [1]{``#1''}%
\providecommand \bibnamefont  [1]{#1}%
\providecommand \bibfnamefont [1]{#1}%
\providecommand \citenamefont [1]{#1}%
\providecommand \href@noop [0]{\@secondoftwo}%
\providecommand \href [0]{\begingroup \@sanitize@url \@href}%
\providecommand \@href[1]{\@@startlink{#1}\@@href}%
\providecommand \@@href[1]{\endgroup#1\@@endlink}%
\providecommand \@sanitize@url [0]{\catcode `\\12\catcode `\$12\catcode
  `\&12\catcode `\#12\catcode `\^12\catcode `\_12\catcode `\%12\relax}%
\providecommand \@@startlink[1]{}%
\providecommand \@@endlink[0]{}%
\providecommand \url  [0]{\begingroup\@sanitize@url \@url }%
\providecommand \@url [1]{\endgroup\@href {#1}{\urlprefix }}%
\providecommand \urlprefix  [0]{URL }%
\providecommand \Eprint [0]{\href }%
\providecommand \doibase [0]{http://dx.doi.org/}%
\providecommand \selectlanguage [0]{\@gobble}%
\providecommand \bibinfo  [0]{\@secondoftwo}%
\providecommand \bibfield  [0]{\@secondoftwo}%
\providecommand \translation [1]{[#1]}%
\providecommand \BibitemOpen [0]{}%
\providecommand \bibitemStop [0]{}%
\providecommand \bibitemNoStop [0]{.\EOS\space}%
\providecommand \EOS [0]{\spacefactor3000\relax}%
\providecommand \BibitemShut  [1]{\csname bibitem#1\endcsname}%
\let\auto@bib@innerbib\@empty
\bibitem [{\citenamefont {Berger}\ \emph {et~al.}(2021)\citenamefont {Berger},
  \citenamefont {Rammelm\"uller}, \citenamefont {Loheac}, \citenamefont
  {Ehmann}, \citenamefont {Braun},\ and\ \citenamefont
  {Drut}}]{Berger:2019odf}%
  \BibitemOpen
  \bibfield  {author} {\bibinfo {author} {\bibfnamefont {C.~E.}\ \bibnamefont
  {Berger}}, \bibinfo {author} {\bibfnamefont {L.}~\bibnamefont
  {Rammelm\"uller}}, \bibinfo {author} {\bibfnamefont {A.~C.}\ \bibnamefont
  {Loheac}}, \bibinfo {author} {\bibfnamefont {F.}~\bibnamefont {Ehmann}},
  \bibinfo {author} {\bibfnamefont {J.}~\bibnamefont {Braun}}, \ and\ \bibinfo
  {author} {\bibfnamefont {J.~E.}\ \bibnamefont {Drut}},\ }\href {\doibase
  10.1016/j.physrep.2020.09.002} {\bibfield  {journal} {\bibinfo  {journal}
  {Phys. Rept.}\ }\textbf {\bibinfo {volume} {892}},\ \bibinfo {pages} {1}
  (\bibinfo {year} {2021})},\ \Eprint {http://arxiv.org/abs/1907.10183}
  {arXiv:1907.10183 [cond-mat.quant-gas]} \BibitemShut {NoStop}%
\bibitem [{\citenamefont {Attanasio}\ \emph {et~al.}(2020)\citenamefont
  {Attanasio}, \citenamefont {J\"ager},\ and\ \citenamefont
  {Ziegler}}]{Attanasio:2020spv}%
  \BibitemOpen
  \bibfield  {author} {\bibinfo {author} {\bibfnamefont {F.}~\bibnamefont
  {Attanasio}}, \bibinfo {author} {\bibfnamefont {B.}~\bibnamefont {J\"ager}},
  \ and\ \bibinfo {author} {\bibfnamefont {F.~P.~G.}\ \bibnamefont {Ziegler}},\
  }\href {\doibase 10.1140/epja/s10050-020-00256-z} {\bibfield  {journal}
  {\bibinfo  {journal} {Eur. Phys. J. A}\ }\textbf {\bibinfo {volume} {56}},\
  \bibinfo {pages} {251} (\bibinfo {year} {2020})},\ \Eprint
  {http://arxiv.org/abs/2006.00476} {arXiv:2006.00476 [hep-lat]} \BibitemShut
  {NoStop}%
\bibitem [{\citenamefont {Alexandru}\ \emph {et~al.}(2022)\citenamefont
  {Alexandru}, \citenamefont {Basar}, \citenamefont {Bedaque},\ and\
  \citenamefont {Warrington}}]{Alexandru:2020wrj}%
  \BibitemOpen
  \bibfield  {author} {\bibinfo {author} {\bibfnamefont {A.}~\bibnamefont
  {Alexandru}}, \bibinfo {author} {\bibfnamefont {G.}~\bibnamefont {Basar}},
  \bibinfo {author} {\bibfnamefont {P.~F.}\ \bibnamefont {Bedaque}}, \ and\
  \bibinfo {author} {\bibfnamefont {N.~C.}\ \bibnamefont {Warrington}},\ }\href
  {\doibase 10.1103/RevModPhys.94.015006} {\bibfield  {journal} {\bibinfo
  {journal} {Rev. Mod. Phys.}\ }\textbf {\bibinfo {volume} {94}},\ \bibinfo
  {pages} {015006} (\bibinfo {year} {2022})},\ \Eprint
  {http://arxiv.org/abs/2007.05436} {arXiv:2007.05436 [hep-lat]} \BibitemShut
  {NoStop}%
\bibitem [{\citenamefont {de~Forcrand}\ and\ \citenamefont
  {Philipsen}(2002)}]{deForcrand:2002hgr}%
  \BibitemOpen
  \bibfield  {author} {\bibinfo {author} {\bibfnamefont {P.}~\bibnamefont
  {de~Forcrand}}\ and\ \bibinfo {author} {\bibfnamefont {O.}~\bibnamefont
  {Philipsen}},\ }\href {\doibase 10.1016/S0550-3213(02)00626-0} {\bibfield
  {journal} {\bibinfo  {journal} {Nucl. Phys. B}\ }\textbf {\bibinfo {volume}
  {642}},\ \bibinfo {pages} {290} (\bibinfo {year} {2002})},\ \Eprint
  {http://arxiv.org/abs/hep-lat/0205016} {arXiv:hep-lat/0205016} \BibitemShut
  {NoStop}%
\bibitem [{\citenamefont {D'Elia}\ and\ \citenamefont
  {Lombardo}(2003)}]{DElia:2002tig}%
  \BibitemOpen
  \bibfield  {author} {\bibinfo {author} {\bibfnamefont {M.}~\bibnamefont
  {D'Elia}}\ and\ \bibinfo {author} {\bibfnamefont {M.-P.}\ \bibnamefont
  {Lombardo}},\ }\href {\doibase 10.1103/PhysRevD.67.014505} {\bibfield
  {journal} {\bibinfo  {journal} {Phys. Rev. D}\ }\textbf {\bibinfo {volume}
  {67}},\ \bibinfo {pages} {014505} (\bibinfo {year} {2003})},\ \Eprint
  {http://arxiv.org/abs/hep-lat/0209146} {arXiv:hep-lat/0209146} \BibitemShut
  {NoStop}%
\bibitem [{\citenamefont {Dimopoulos}\ \emph {et~al.}(2022)\citenamefont
  {Dimopoulos}, \citenamefont {Dini}, \citenamefont {Di~Renzo}, \citenamefont
  {Goswami}, \citenamefont {Nicotra}, \citenamefont {Schmidt}, \citenamefont
  {Singh}, \citenamefont {Zambello},\ and\ \citenamefont
  {Ziesch\'e}}]{Dimopoulos:2021vrk}%
  \BibitemOpen
  \bibfield  {author} {\bibinfo {author} {\bibfnamefont {P.}~\bibnamefont
  {Dimopoulos}}, \bibinfo {author} {\bibfnamefont {L.}~\bibnamefont {Dini}},
  \bibinfo {author} {\bibfnamefont {F.}~\bibnamefont {Di~Renzo}}, \bibinfo
  {author} {\bibfnamefont {J.}~\bibnamefont {Goswami}}, \bibinfo {author}
  {\bibfnamefont {G.}~\bibnamefont {Nicotra}}, \bibinfo {author} {\bibfnamefont
  {C.}~\bibnamefont {Schmidt}}, \bibinfo {author} {\bibfnamefont
  {S.}~\bibnamefont {Singh}}, \bibinfo {author} {\bibfnamefont
  {K.}~\bibnamefont {Zambello}}, \ and\ \bibinfo {author} {\bibfnamefont
  {F.}~\bibnamefont {Ziesch\'e}},\ }\href {\doibase
  10.1103/PhysRevD.105.034513} {\bibfield  {journal} {\bibinfo  {journal}
  {Phys. Rev. D}\ }\textbf {\bibinfo {volume} {105}},\ \bibinfo {pages}
  {034513} (\bibinfo {year} {2022})},\ \Eprint
  {http://arxiv.org/abs/2110.15933} {arXiv:2110.15933 [hep-lat]} \BibitemShut
  {NoStop}%
\bibitem [{\citenamefont {Roberge}\ and\ \citenamefont
  {Weiss}(1986)}]{Roberge:1986mm}%
  \BibitemOpen
  \bibfield  {author} {\bibinfo {author} {\bibfnamefont {A.}~\bibnamefont
  {Roberge}}\ and\ \bibinfo {author} {\bibfnamefont {N.}~\bibnamefont
  {Weiss}},\ }\href {\doibase 10.1016/0550-3213(86)90582-1} {\bibfield
  {journal} {\bibinfo  {journal} {Nucl. Phys. B}\ }\textbf {\bibinfo {volume}
  {275}},\ \bibinfo {pages} {734} (\bibinfo {year} {1986})}\BibitemShut
  {NoStop}%
\bibitem [{\citenamefont {Bornyakov}\ \emph {et~al.}(2023)\citenamefont
  {Bornyakov}, \citenamefont {Gerasimeniuk}, \citenamefont {Goy}, \citenamefont
  {Korneev}, \citenamefont {Molochkov}, \citenamefont {Nakamura},\ and\
  \citenamefont {Rogalyov}}]{Bornyakov:2022blw}%
  \BibitemOpen
  \bibfield  {author} {\bibinfo {author} {\bibfnamefont {V.~G.}\ \bibnamefont
  {Bornyakov}}, \bibinfo {author} {\bibfnamefont {N.~V.}\ \bibnamefont
  {Gerasimeniuk}}, \bibinfo {author} {\bibfnamefont {V.~A.}\ \bibnamefont
  {Goy}}, \bibinfo {author} {\bibfnamefont {A.~A.}\ \bibnamefont {Korneev}},
  \bibinfo {author} {\bibfnamefont {A.~V.}\ \bibnamefont {Molochkov}}, \bibinfo
  {author} {\bibfnamefont {A.}~\bibnamefont {Nakamura}}, \ and\ \bibinfo
  {author} {\bibfnamefont {R.~N.}\ \bibnamefont {Rogalyov}},\ }\href {\doibase
  10.1103/PhysRevD.107.014508} {\bibfield  {journal} {\bibinfo  {journal}
  {Phys. Rev. D}\ }\textbf {\bibinfo {volume} {107}},\ \bibinfo {pages}
  {014508} (\bibinfo {year} {2023})},\ \Eprint
  {http://arxiv.org/abs/2203.06159} {arXiv:2203.06159 [hep-lat]} \BibitemShut
  {NoStop}%
\bibitem [{\citenamefont {Kortman}\ and\ \citenamefont
  {Griffiths}(1971)}]{Kortman:1971zz}%
  \BibitemOpen
  \bibfield  {author} {\bibinfo {author} {\bibfnamefont {P.~J.}\ \bibnamefont
  {Kortman}}\ and\ \bibinfo {author} {\bibfnamefont {R.~B.}\ \bibnamefont
  {Griffiths}},\ }\href {\doibase 10.1103/PhysRevLett.27.1439} {\bibfield
  {journal} {\bibinfo  {journal} {Phys. Rev. Lett.}\ }\textbf {\bibinfo
  {volume} {27}},\ \bibinfo {pages} {1439} (\bibinfo {year}
  {1971})}\BibitemShut {NoStop}%
\bibitem [{\citenamefont {Fisher}(1978)}]{Fisher:1978pf}%
  \BibitemOpen
  \bibfield  {author} {\bibinfo {author} {\bibfnamefont {M.~E.}\ \bibnamefont
  {Fisher}},\ }\href {\doibase 10.1103/PhysRevLett.40.1610} {\bibfield
  {journal} {\bibinfo  {journal} {Phys. Rev. Lett.}\ }\textbf {\bibinfo
  {volume} {40}},\ \bibinfo {pages} {1610} (\bibinfo {year}
  {1978})}\BibitemShut {NoStop}%
\bibitem [{\citenamefont {Lee}\ and\ \citenamefont {Yang}(1952)}]{Lee:1952ig}%
  \BibitemOpen
  \bibfield  {author} {\bibinfo {author} {\bibfnamefont {T.~D.}\ \bibnamefont
  {Lee}}\ and\ \bibinfo {author} {\bibfnamefont {C.-N.}\ \bibnamefont {Yang}},\
  }\href {\doibase 10.1103/PhysRev.87.410} {\bibfield  {journal} {\bibinfo
  {journal} {Phys. Rev.}\ }\textbf {\bibinfo {volume} {87}},\ \bibinfo {pages}
  {410} (\bibinfo {year} {1952})}\BibitemShut {NoStop}%
\bibitem [{\citenamefont {Yang}\ and\ \citenamefont {Lee}(1952)}]{Yang:1952be}%
  \BibitemOpen
  \bibfield  {author} {\bibinfo {author} {\bibfnamefont {C.-N.}\ \bibnamefont
  {Yang}}\ and\ \bibinfo {author} {\bibfnamefont {T.~D.}\ \bibnamefont {Lee}},\
  }\href {\doibase 10.1103/PhysRev.87.404} {\bibfield  {journal} {\bibinfo
  {journal} {Phys. Rev.}\ }\textbf {\bibinfo {volume} {87}},\ \bibinfo {pages}
  {404} (\bibinfo {year} {1952})}\BibitemShut {NoStop}%
\bibitem [{\citenamefont {Itzykson}\ \emph {et~al.}(1983)\citenamefont
  {Itzykson}, \citenamefont {Pearson},\ and\ \citenamefont
  {Zuber}}]{Itzykson:1983gb}%
  \BibitemOpen
  \bibfield  {author} {\bibinfo {author} {\bibfnamefont {C.}~\bibnamefont
  {Itzykson}}, \bibinfo {author} {\bibfnamefont {R.~B.}\ \bibnamefont
  {Pearson}}, \ and\ \bibinfo {author} {\bibfnamefont {J.~B.}\ \bibnamefont
  {Zuber}},\ }\href {\doibase 10.1016/0550-3213(83)90499-6} {\bibfield
  {journal} {\bibinfo  {journal} {Nucl. Phys. B}\ }\textbf {\bibinfo {volume}
  {220}},\ \bibinfo {pages} {415} (\bibinfo {year} {1983})}\BibitemShut
  {NoStop}%
\bibitem [{\citenamefont {Stephanov}(2006)}]{Stephanov:2006dn}%
  \BibitemOpen
  \bibfield  {author} {\bibinfo {author} {\bibfnamefont {M.~A.}\ \bibnamefont
  {Stephanov}},\ }\href {\doibase 10.1103/PhysRevD.73.094508} {\bibfield
  {journal} {\bibinfo  {journal} {Phys. Rev. D}\ }\textbf {\bibinfo {volume}
  {73}},\ \bibinfo {pages} {094508} (\bibinfo {year} {2006})},\ \Eprint
  {http://arxiv.org/abs/hep-lat/0603014} {arXiv:hep-lat/0603014} \BibitemShut
  {NoStop}%
\bibitem [{\citenamefont {Gliozzi}\ and\ \citenamefont
  {Rago}(2014)}]{Gliozzi:2014jsa}%
  \BibitemOpen
  \bibfield  {author} {\bibinfo {author} {\bibfnamefont {F.}~\bibnamefont
  {Gliozzi}}\ and\ \bibinfo {author} {\bibfnamefont {A.}~\bibnamefont {Rago}},\
  }\href {\doibase 10.1007/JHEP10(2014)042} {\bibfield  {journal} {\bibinfo
  {journal} {JHEP}\ }\textbf {\bibinfo {volume} {10}},\ \bibinfo {pages} {042}
  (\bibinfo {year} {2014})},\ \Eprint {http://arxiv.org/abs/1403.6003}
  {arXiv:1403.6003 [hep-th]} \BibitemShut {NoStop}%
\bibitem [{\citenamefont {Alm\'asi}\ \emph {et~al.}(2019)\citenamefont
  {Alm\'asi}, \citenamefont {Friman}, \citenamefont {Morita},\ and\
  \citenamefont {Redlich}}]{Almasi:2019bvl}%
  \BibitemOpen
  \bibfield  {author} {\bibinfo {author} {\bibfnamefont {G.~A.}\ \bibnamefont
  {Alm\'asi}}, \bibinfo {author} {\bibfnamefont {B.}~\bibnamefont {Friman}},
  \bibinfo {author} {\bibfnamefont {K.}~\bibnamefont {Morita}}, \ and\ \bibinfo
  {author} {\bibfnamefont {K.}~\bibnamefont {Redlich}},\ }\href {\doibase
  10.1016/j.physletb.2019.04.023} {\bibfield  {journal} {\bibinfo  {journal}
  {Phys. Lett. B}\ }\textbf {\bibinfo {volume} {793}},\ \bibinfo {pages} {19}
  (\bibinfo {year} {2019})},\ \Eprint {http://arxiv.org/abs/1902.05457}
  {arXiv:1902.05457 [hep-ph]} \BibitemShut {NoStop}%
\bibitem [{\citenamefont {Vovchenko}\ \emph {et~al.}(2018)\citenamefont
  {Vovchenko}, \citenamefont {Steinheimer}, \citenamefont {Philipsen},\ and\
  \citenamefont {Stoecker}}]{Vovchenko:2017gkg}%
  \BibitemOpen
  \bibfield  {author} {\bibinfo {author} {\bibfnamefont {V.}~\bibnamefont
  {Vovchenko}}, \bibinfo {author} {\bibfnamefont {J.}~\bibnamefont
  {Steinheimer}}, \bibinfo {author} {\bibfnamefont {O.}~\bibnamefont
  {Philipsen}}, \ and\ \bibinfo {author} {\bibfnamefont {H.}~\bibnamefont
  {Stoecker}},\ }\href {\doibase 10.1103/PhysRevD.97.114030} {\bibfield
  {journal} {\bibinfo  {journal} {Phys. Rev. D}\ }\textbf {\bibinfo {volume}
  {97}},\ \bibinfo {pages} {114030} (\bibinfo {year} {2018})},\ \Eprint
  {http://arxiv.org/abs/1711.01261} {arXiv:1711.01261 [hep-ph]} \BibitemShut
  {NoStop}%
\bibitem [{\citenamefont {Vovchenko}\ \emph {et~al.}(2017)\citenamefont
  {Vovchenko}, \citenamefont {Pasztor}, \citenamefont {Fodor}, \citenamefont
  {Katz},\ and\ \citenamefont {Stoecker}}]{Vovchenko:2017xad}%
  \BibitemOpen
  \bibfield  {author} {\bibinfo {author} {\bibfnamefont {V.}~\bibnamefont
  {Vovchenko}}, \bibinfo {author} {\bibfnamefont {A.}~\bibnamefont {Pasztor}},
  \bibinfo {author} {\bibfnamefont {Z.}~\bibnamefont {Fodor}}, \bibinfo
  {author} {\bibfnamefont {S.~D.}\ \bibnamefont {Katz}}, \ and\ \bibinfo
  {author} {\bibfnamefont {H.}~\bibnamefont {Stoecker}},\ }\href {\doibase
  10.1016/j.physletb.2017.10.042} {\bibfield  {journal} {\bibinfo  {journal}
  {Phys. Lett. B}\ }\textbf {\bibinfo {volume} {775}},\ \bibinfo {pages} {71}
  (\bibinfo {year} {2017})},\ \Eprint {http://arxiv.org/abs/1708.02852}
  {arXiv:1708.02852 [hep-ph]} \BibitemShut {NoStop}%
\bibitem [{\citenamefont {Almasi}\ \emph {et~al.}(2019)\citenamefont {Almasi},
  \citenamefont {Friman}, \citenamefont {Morita}, \citenamefont {Lo},\ and\
  \citenamefont {Redlich}}]{Almasi:2018lok}%
  \BibitemOpen
  \bibfield  {author} {\bibinfo {author} {\bibfnamefont {G.~A.}\ \bibnamefont
  {Almasi}}, \bibinfo {author} {\bibfnamefont {B.}~\bibnamefont {Friman}},
  \bibinfo {author} {\bibfnamefont {K.}~\bibnamefont {Morita}}, \bibinfo
  {author} {\bibfnamefont {P.~M.}\ \bibnamefont {Lo}}, \ and\ \bibinfo {author}
  {\bibfnamefont {K.}~\bibnamefont {Redlich}},\ }\href {\doibase
  10.1103/PhysRevD.100.016016} {\bibfield  {journal} {\bibinfo  {journal}
  {Phys. Rev. D}\ }\textbf {\bibinfo {volume} {100}},\ \bibinfo {pages}
  {016016} (\bibinfo {year} {2019})},\ \Eprint
  {http://arxiv.org/abs/1805.04441} {arXiv:1805.04441 [hep-ph]} \BibitemShut
  {NoStop}%
\bibitem [{\citenamefont {Bornyakov}\ \emph {et~al.}(2017)\citenamefont
  {Bornyakov}, \citenamefont {Boyda}, \citenamefont {Goy}, \citenamefont
  {Molochkov}, \citenamefont {Nakamura}, \citenamefont {Nikolaev},\ and\
  \citenamefont {Zakharov}}]{Bornyakov:2016wld}%
  \BibitemOpen
  \bibfield  {author} {\bibinfo {author} {\bibfnamefont {V.~G.}\ \bibnamefont
  {Bornyakov}}, \bibinfo {author} {\bibfnamefont {D.~L.}\ \bibnamefont
  {Boyda}}, \bibinfo {author} {\bibfnamefont {V.~A.}\ \bibnamefont {Goy}},
  \bibinfo {author} {\bibfnamefont {A.~V.}\ \bibnamefont {Molochkov}}, \bibinfo
  {author} {\bibfnamefont {A.}~\bibnamefont {Nakamura}}, \bibinfo {author}
  {\bibfnamefont {A.~A.}\ \bibnamefont {Nikolaev}}, \ and\ \bibinfo {author}
  {\bibfnamefont {V.~I.}\ \bibnamefont {Zakharov}},\ }\href {\doibase
  10.1103/PhysRevD.95.094506} {\bibfield  {journal} {\bibinfo  {journal} {Phys.
  Rev. D}\ }\textbf {\bibinfo {volume} {95}},\ \bibinfo {pages} {094506}
  (\bibinfo {year} {2017})},\ \Eprint {http://arxiv.org/abs/1611.04229}
  {arXiv:1611.04229 [hep-lat]} \BibitemShut {NoStop}%
\bibitem [{\citenamefont {Schmidt}(2023)}]{Schmidt:2023jcv}%
  \BibitemOpen
  \bibfield  {author} {\bibinfo {author} {\bibfnamefont {C.}~\bibnamefont
  {Schmidt}},\ }\href {\doibase 10.22323/1.430.0159} {\bibfield  {journal}
  {\bibinfo  {journal} {PoS}\ }\textbf {\bibinfo {volume} {LATTICE2022}},\
  \bibinfo {pages} {159} (\bibinfo {year} {2023})},\ \Eprint
  {http://arxiv.org/abs/2301.04978} {arXiv:2301.04978 [hep-lat]} \BibitemShut
  {NoStop}%
\bibitem [{\citenamefont {Borinsky}\ \emph {et~al.}(2021)\citenamefont
  {Borinsky}, \citenamefont {Gracey}, \citenamefont {Kompaniets},\ and\
  \citenamefont {Schnetz}}]{Borinsky:2021jdb}%
  \BibitemOpen
  \bibfield  {author} {\bibinfo {author} {\bibfnamefont {M.}~\bibnamefont
  {Borinsky}}, \bibinfo {author} {\bibfnamefont {J.~A.}\ \bibnamefont
  {Gracey}}, \bibinfo {author} {\bibfnamefont {M.~V.}\ \bibnamefont
  {Kompaniets}}, \ and\ \bibinfo {author} {\bibfnamefont {O.}~\bibnamefont
  {Schnetz}},\ }\href {\doibase 10.1103/PhysRevD.103.116024} {\bibfield
  {journal} {\bibinfo  {journal} {Phys. Rev. D}\ }\textbf {\bibinfo {volume}
  {103}},\ \bibinfo {pages} {116024} (\bibinfo {year} {2021})},\ \Eprint
  {http://arxiv.org/abs/2103.16224} {arXiv:2103.16224 [hep-th]} \BibitemShut
  {NoStop}%
\bibitem [{\citenamefont {Connelly}\ \emph {et~al.}(2020)\citenamefont
  {Connelly}, \citenamefont {Johnson}, \citenamefont {Rennecke},\ and\
  \citenamefont {Skokov}}]{Connelly:2020gwa}%
  \BibitemOpen
  \bibfield  {author} {\bibinfo {author} {\bibfnamefont {A.}~\bibnamefont
  {Connelly}}, \bibinfo {author} {\bibfnamefont {G.}~\bibnamefont {Johnson}},
  \bibinfo {author} {\bibfnamefont {F.}~\bibnamefont {Rennecke}}, \ and\
  \bibinfo {author} {\bibfnamefont {V.}~\bibnamefont {Skokov}},\ }\href
  {\doibase 10.1103/PhysRevLett.125.191602} {\bibfield  {journal} {\bibinfo
  {journal} {Phys. Rev. Lett.}\ }\textbf {\bibinfo {volume} {125}},\ \bibinfo
  {pages} {191602} (\bibinfo {year} {2020})},\ \Eprint
  {http://arxiv.org/abs/2006.12541} {arXiv:2006.12541 [cond-mat.stat-mech]}
  \BibitemShut {NoStop}%
\bibitem [{\citenamefont {Rennecke}\ and\ \citenamefont
  {Skokov}(2022)}]{Rennecke:2022ohx}%
  \BibitemOpen
  \bibfield  {author} {\bibinfo {author} {\bibfnamefont {F.}~\bibnamefont
  {Rennecke}}\ and\ \bibinfo {author} {\bibfnamefont {V.~V.}\ \bibnamefont
  {Skokov}},\ }\href {\doibase 10.1016/j.aop.2022.169010} {\bibfield  {journal}
  {\bibinfo  {journal} {Annals Phys.}\ }\textbf {\bibinfo {volume} {444}},\
  \bibinfo {pages} {169010} (\bibinfo {year} {2022})},\ \Eprint
  {http://arxiv.org/abs/2203.16651} {arXiv:2203.16651 [hep-ph]} \BibitemShut
  {NoStop}%
\bibitem [{\citenamefont {Johnson}\ \emph {et~al.}(2023)\citenamefont
  {Johnson}, \citenamefont {Rennecke},\ and\ \citenamefont
  {Skokov}}]{Johnson:2022cqv}%
  \BibitemOpen
  \bibfield  {author} {\bibinfo {author} {\bibfnamefont {G.}~\bibnamefont
  {Johnson}}, \bibinfo {author} {\bibfnamefont {F.}~\bibnamefont {Rennecke}}, \
  and\ \bibinfo {author} {\bibfnamefont {V.~V.}\ \bibnamefont {Skokov}},\
  }\href {\doibase 10.1103/PhysRevD.107.116013} {\bibfield  {journal} {\bibinfo
   {journal} {Phys. Rev. D}\ }\textbf {\bibinfo {volume} {107}},\ \bibinfo
  {pages} {116013} (\bibinfo {year} {2023})},\ \Eprint
  {http://arxiv.org/abs/2211.00710} {arXiv:2211.00710 [hep-ph]} \BibitemShut
  {NoStop}%
\bibitem [{\citenamefont {Clarke}\ \emph {et~al.}(2023)\citenamefont {Clarke},
  \citenamefont {Zambello}, \citenamefont {Dimopoulos}, \citenamefont
  {Di~Renzo}, \citenamefont {Goswami}, \citenamefont {Nicotra}, \citenamefont
  {Schmidt},\ and\ \citenamefont {Singh}}]{Clarke:2023noy}%
  \BibitemOpen
  \bibfield  {author} {\bibinfo {author} {\bibfnamefont {D.~A.}\ \bibnamefont
  {Clarke}}, \bibinfo {author} {\bibfnamefont {K.}~\bibnamefont {Zambello}},
  \bibinfo {author} {\bibfnamefont {P.}~\bibnamefont {Dimopoulos}}, \bibinfo
  {author} {\bibfnamefont {F.}~\bibnamefont {Di~Renzo}}, \bibinfo {author}
  {\bibfnamefont {J.}~\bibnamefont {Goswami}}, \bibinfo {author} {\bibfnamefont
  {G.}~\bibnamefont {Nicotra}}, \bibinfo {author} {\bibfnamefont
  {C.}~\bibnamefont {Schmidt}}, \ and\ \bibinfo {author} {\bibfnamefont
  {S.}~\bibnamefont {Singh}},\ }\href {\doibase 10.22323/1.430.0164} {\bibfield
   {journal} {\bibinfo  {journal} {PoS}\ }\textbf {\bibinfo {volume}
  {LATTICE2022}},\ \bibinfo {pages} {164} (\bibinfo {year} {2023})},\ \Eprint
  {http://arxiv.org/abs/2301.03952} {arXiv:2301.03952 [hep-lat]} \BibitemShut
  {NoStop}%
\bibitem [{\citenamefont {Mukherjee}\ and\ \citenamefont
  {Skokov}(2021)}]{Mukherjee:2019eou}%
  \BibitemOpen
  \bibfield  {author} {\bibinfo {author} {\bibfnamefont {S.}~\bibnamefont
  {Mukherjee}}\ and\ \bibinfo {author} {\bibfnamefont {V.}~\bibnamefont
  {Skokov}},\ }\href {\doibase 10.1103/PhysRevD.103.L071501} {\bibfield
  {journal} {\bibinfo  {journal} {Phys. Rev. D}\ }\textbf {\bibinfo {volume}
  {103}},\ \bibinfo {pages} {L071501} (\bibinfo {year} {2021})},\ \Eprint
  {http://arxiv.org/abs/1909.04639} {arXiv:1909.04639 [hep-ph]} \BibitemShut
  {NoStop}%
\bibitem [{\citenamefont {Mukherjee}\ \emph {et~al.}(2022)\citenamefont
  {Mukherjee}, \citenamefont {Rennecke},\ and\ \citenamefont
  {Skokov}}]{Mukherjee:2021tyg}%
  \BibitemOpen
  \bibfield  {author} {\bibinfo {author} {\bibfnamefont {S.}~\bibnamefont
  {Mukherjee}}, \bibinfo {author} {\bibfnamefont {F.}~\bibnamefont {Rennecke}},
  \ and\ \bibinfo {author} {\bibfnamefont {V.~V.}\ \bibnamefont {Skokov}},\
  }\href {\doibase 10.1103/PhysRevD.105.014026} {\bibfield  {journal} {\bibinfo
   {journal} {Phys. Rev. D}\ }\textbf {\bibinfo {volume} {105}},\ \bibinfo
  {pages} {014026} (\bibinfo {year} {2022})},\ \Eprint
  {http://arxiv.org/abs/2110.02241} {arXiv:2110.02241 [hep-ph]} \BibitemShut
  {NoStop}%
\bibitem [{\citenamefont {Skokov}\ \emph
  {et~al.}(2010{\natexlab{a}})\citenamefont {Skokov}, \citenamefont {Friman},
  \citenamefont {Nakano}, \citenamefont {Redlich},\ and\ \citenamefont
  {Schaefer}}]{Skokov:2010sf}%
  \BibitemOpen
  \bibfield  {author} {\bibinfo {author} {\bibfnamefont {V.}~\bibnamefont
  {Skokov}}, \bibinfo {author} {\bibfnamefont {B.}~\bibnamefont {Friman}},
  \bibinfo {author} {\bibfnamefont {E.}~\bibnamefont {Nakano}}, \bibinfo
  {author} {\bibfnamefont {K.}~\bibnamefont {Redlich}}, \ and\ \bibinfo
  {author} {\bibfnamefont {B.~J.}\ \bibnamefont {Schaefer}},\ }\href {\doibase
  10.1103/PhysRevD.82.034029} {\bibfield  {journal} {\bibinfo  {journal} {Phys.
  Rev. D}\ }\textbf {\bibinfo {volume} {82}},\ \bibinfo {pages} {034029}
  (\bibinfo {year} {2010}{\natexlab{a}})},\ \Eprint
  {http://arxiv.org/abs/1005.3166} {arXiv:1005.3166 [hep-ph]} \BibitemShut
  {NoStop}%
\bibitem [{\citenamefont {Skokov}\ \emph
  {et~al.}(2010{\natexlab{b}})\citenamefont {Skokov}, \citenamefont {Stokic},
  \citenamefont {Friman},\ and\ \citenamefont {Redlich}}]{Skokov:2010wb}%
  \BibitemOpen
  \bibfield  {author} {\bibinfo {author} {\bibfnamefont {V.}~\bibnamefont
  {Skokov}}, \bibinfo {author} {\bibfnamefont {B.}~\bibnamefont {Stokic}},
  \bibinfo {author} {\bibfnamefont {B.}~\bibnamefont {Friman}}, \ and\ \bibinfo
  {author} {\bibfnamefont {K.}~\bibnamefont {Redlich}},\ }\href {\doibase
  10.1103/PhysRevC.82.015206} {\bibfield  {journal} {\bibinfo  {journal} {Phys.
  Rev. C}\ }\textbf {\bibinfo {volume} {82}},\ \bibinfo {pages} {015206}
  (\bibinfo {year} {2010}{\natexlab{b}})},\ \Eprint
  {http://arxiv.org/abs/1004.2665} {arXiv:1004.2665 [hep-ph]} \BibitemShut
  {NoStop}%
\bibitem [{\citenamefont {Deaño}\ \emph {et~al.}(2017)\citenamefont {Deaño},
  \citenamefont {Huybrechs},\ and\ \citenamefont
  {Iserles}}]{doi:10.1137/1.9781611975123}%
  \BibitemOpen
  \bibfield  {author} {\bibinfo {author} {\bibfnamefont {A.}~\bibnamefont
  {Deaño}}, \bibinfo {author} {\bibfnamefont {D.}~\bibnamefont {Huybrechs}}, \
  and\ \bibinfo {author} {\bibfnamefont {A.}~\bibnamefont {Iserles}},\ }\href
  {\doibase 10.1137/1.9781611975123} {\emph {\bibinfo {title} {Computing Highly
  Oscillatory Integrals}}}\ (\bibinfo  {publisher} {Society for Industrial and
  Applied Mathematics},\ \bibinfo {address} {Philadelphia, PA},\ \bibinfo
  {year} {2017})\ \Eprint
  {http://arxiv.org/abs/https://epubs.siam.org/doi/pdf/10.1137/1.9781611975123}
  {https://epubs.siam.org/doi/pdf/10.1137/1.9781611975123} \BibitemShut
  {NoStop}%
\end{thebibliography}%

\end{document}